%% file: Script_part1.tex
\documentclass[twoside,openright,a4paper,12pt]{article}

\newif\iffull
\fullfalse 

\usepackage[left=2cm,right=2cm,top=1cm,bottom=1cm, includeheadfoot]{geometry}
\usepackage[english]{babel}
\usepackage[utf8]{inputenc}
\usepackage{textcomp}
\usepackage[T1]{fontenc}
\usepackage{lmodern}
\usepackage{amsmath, amsfonts, amssymb, cancel, amsthm}

\usepackage{graphicx}

\usepackage{shadethm}

\usepackage{enumerate}

\newshadetheorem{mydef}{Definition}[section]
\newshadetheorem{example}{Example}[section]
\newshadetheorem{theo}{Theorem}[section]
\newshadetheorem{coro}{Corollary}[section]
\newshadetheorem{lemma}{Lemma}[section]

\newcommand{\mathscr}[1]{\mathcal{#1}}
 
\newenvironment{ex}[1][]{
\subsubsection{Example: #1} 
}{}

\DeclareMathOperator{\Tr}{Tr}
\DeclareMathOperator{\tr}{Tr}
 
\newcommand{\funs}{L}
\newcommand{\qcmap}{\mathcal X}
\newcommand{\mc}[1]{\mathcal #1}
 
\newcommand{\re}{\textrm Re\,}
\newcommand{\im}{\textrm Im\,}
\newcommand{\one}{{\bf 1}}
\newcommand{\identity}{\one} 

\newcommand{\channel}{\mathcal{E}}
\newcommand{\algebra}{\mathscr{A}}
\newcommand{\hilbert}{\mathcal{H}}
\newcommand{\tensor}{\otimes}
\newcommand{\el}{\; \epsilon \;}

\newcommand{\ket}[1]{|{#1}\rangle} 
\newcommand{\expect}[1]{\langle{#1}\rangle} 
\newcommand{\bra}[1]{\langle{#1}|} 
\newcommand{\braket}[2]{\langle{#1|#2}\rangle}
\newcommand{\proj}[1]{\ket{#1}\bra{#1}}
\newcommand{\ketbra}[2]{\ket{#1}\bra{#2}}
\newcommand{\expectop}[2]{\langle{#1}|#2|{#1}\rangle}

\newcommand{\av}[1]{\langle{#1}\rangle}
\newcommand{\ave}[1]{\langle{#1}\rangle}

\renewcommand{\emph}[1]{\textit{#1}}   
\let\oldqed\qed
\renewcommand{\qed}{\oldqed \vspace{0.5cm}}

\begin{document}
	\thispagestyle{empty}
	\begin{figure}
		\begin{flushright}
			\includegraphics[width=0.3\textwidth]{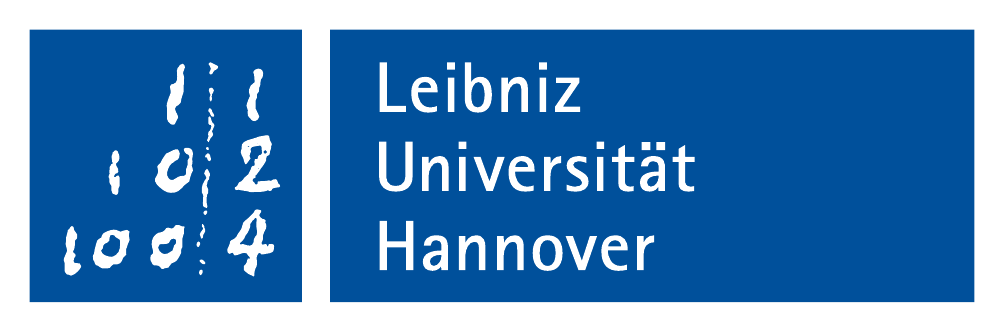}
		\end{flushright}
	\end{figure}
	\vspace{2cm}
	\begin{center}
		\textbf{\huge{Algebraic approach to quantum theory: \\a finite-dimensional guide}}\\
		\vspace{2cm}
		C\'edric B\'eny\\Florian Richter
		\vspace{0.2cm}
	\end{center}
\newpage
\thispagestyle{empty}
\tableofcontents 

\newpage

\pagenumbering{arabic}
\setcounter{page}{1}

\section{Introduction}

This document is meant as a pedagogical introduction to the modern language used to talk about quantum theories, especially in the field of quantum information. It assumes that the reader has taken a first traditional course on quantum mechanics, and is familiar with the concept of Hilbert space and elementary linear algebra. 

As in the popular textbook on quantum information by Nielsen and Chuang~\cite{nielsen}, we introduce the generalised concept of states (density matrices), observables (POVMs) and transformations (channels), but we also go further and characterise these structures from an algebraic standpoint, which provides many useful technical tools, and clarity as to their generality. This approach also makes it manifest that quantum theory is a direct generalisation of probability theory, and provides a unifying formalism for both fields.

Although this algebraic approach dates back, in parts, to John von Neumann, we are not aware of any presentation which focuses on finite-dimensional systems. This simplification allows us to have a self-contained presentation which avoids many of the technicalities inherent to the most general $C^*$-algebraic approach, while being perfectly appropriate for the quantum information literature.

\section{States and effects}
\input{BasicStates}

\input{Observables}

\input{channels}


\newpage
\bibliography{Script}
\bibliographystyle{unsrt}
\end{document}

%% file: BasicStates.tex
\subsection{Basic Quantum Mechanics}

Let us start by reviewing the standard 1930's formulation of quantum mechanics, which is still used in many textbooks. The theory is defined around a Hilbert space $\mathcal{H}$, whose normalised vectors $\ket{\psi}\in\mathcal{H}$ represent \textit ph{states}, or ``wave functions''. The \textit ph{observables} of the theory are represented by the set of self-adjoint operators $A^\dagger = A$ acting on $\mathcal{H}$. 

The operational meaning of the observable has to be ``unpacked'' using the spectral theorem. For instance, if $\mathcal H$ has a finite dimension $d$, it states that there exists an orthonormal basis $\{\ket{i}\}_{i=1}^{d}$, i.e., $\braket{i}{j}=\delta_{i,j}$, such that 
\(
A\ket i = a_i \ket i,
\) 
where $a_i\in\mathbb{R}$ are the eigenvalues of $A$. 

In the simple case where the eigenvalues $a_i$ are all distinct, the operational content of the formalism is summarised in the following statement: when measuring the observable $A$ on a quantum system in state $\ket{\psi}$, the probability of obtaining the $i$th outcome (where the observable takes the value $a_i$) is
$
p_i=\braket{i}{\psi}\braket{\psi}{i}=|\braket{i}{\psi}|^2.
$

More generally, if the operator $A$ has degenerate eigenvalues, it is convenient to write $A\ket{i,k}=a_i\ket{i,k}$, where $k$ labels the degeneracy of eigenvalue $a_i$, and the $a_i$ are all distinct. The probability to obtain the $i$th outcome is then determined by the famous 

\begin{mydef}[{Born Rule}] If the system is in state $\ket \psi$, a measurement of the observable represented by the self-adjoint operator $A$ returns the value $a_i$ with probability $p_i=\sum_k|\braket{i,k}{\psi}|^2$.
\end{mydef}

Normally, one goes on to describe the projection postulate, but we will see that it can actually be derived from the born rule, once properly generalised.

\subsection{Positive operators}
\label{positive:operators}

In the first part of the lecture we are going to generalize this description of quantum mechanics, and in some sense simplify it. 
For this purpose, it will be useful to understand the concept of positivity of operators.


\begin{mydef}
An operator $A$ is called \textbf{positive}, denoted as\ $A\geq 0$, if for all $\ket{\psi}\in\mathcal{H}$, $\bra{\psi}A\ket{\psi}\geq0$. We also write $A\geq B$ if $A-B\geq 0$.
\end{mydef}

In finite dimension, this definition is equivalent to $A$ being self-adjoint and having non-negative eigenvalues, that is \ $A^\dagger =A$, $\forall i: a_i\geq 0$. It is straightforward to see that such an operator implies the defining properties. Let $\{\ket{i}\}$ denote the eigenbasis of $A$:
\[
\bra{\psi}A\ket{\psi}=\sum_{i,j}\overline{c_i}{c_j}\bra{i}a_j\ket{j}=\sum_i|c_i|^2a_i\geq 0.
\] 
For the converse, consider $A$ with $\bra{\psi}A\ket{\psi}\geq0$ for all $\ket{\psi}\in\mathcal{H}$.
To see that this $A$ is self-adjoint, observe that in general,
\begin{equation}
\begin{split}
\bra{\psi + \phi}A\ket{\psi + \phi} 
&= \bra{\psi}A\ket{\psi}+\bra{\phi}A\ket{\phi} + \bra{\psi}A\ket{\phi} + \bra{\phi}A\ket{\psi}.\\
\end{split}
\end{equation}
Since this is real by assumption, and also the first two terms on the right side of the equality are real, we must have
\[
\bra{\psi}A\ket{\phi} + \bra{\phi}A\ket{\psi} = \overline{ \bra{\psi}A\ket{\phi}} + \overline{\bra{\phi}A\ket{\psi}}
\]
for all states $\psi$ and $\phi$, which implies $\im \bra{\psi}A\ket{\phi} = - \im \bra{\phi}A\ket{\psi}$. Replacing $\phi$ by $i \phi$, we also obtain $\re \bra{\psi}A\ket{\phi} = \re \bra{\phi}A\ket{\psi}$. It follows that $\bra{\psi}A\ket{\phi} = \overline{\bra{\phi}A\ket{\psi}}$, and hence $A$ is self-adjoint. The positivity of the eigenvalues follows from the fact that eigenvalues can be considered as expectation values of eigenvectors, i.e.\ $a_i=\bra{i}A\ket{i}\geq 0$.

Another important point to notice is that for any operator $B$, we have $B^\dagger B\geq 0$, because for all $\psi$, $\bra{\psi}B^\dagger B \ket{\psi}=\|B\ket{\psi}\|^2\geq 0$. It is also true that any positive operator $A$ can be written in this form by using, for instance, $B = \sqrt{A}$, defined as having the same eigenvectors as $A$, but the square root of the eigenvectors of $A$. 

Also, we will use the fact that the binary relation $A\geq B$ if $B-A\geq 0$ defines a partial order on the set of all operators. This order is partial because not every pair of operators $A,B$ can be compared. For instance, consider the matrices 
\[
A=\begin{pmatrix} 1 & 0\\0 &0 \end{pmatrix} \quad\text{and}\quad B=\begin{pmatrix} 0 & 0\\0 &1 \end{pmatrix}.
\]
It is easy to see that neither $A-B$ nor $B-A$ are positive, since they both have a negative eigenvalue.

An important set of positive operators are the projectors:
\begin{mydef}
An operator $P$ is a \textbf{projector}, if it satisfies $P^\dagger P=P$ or, equivalently, $P=P^\dagger $ and $P^2=P$. 
\end{mydef}

It follows that the eigenvalues of a projector $P$ are either $0$ or $1$. Indeed, if $P \ket \psi = p \ket \psi$ for some eigenvalue $p$, then also $p \ket \psi = P \ket \psi = P^2 \ket \psi = p^2 \ket \psi$, it follows that $p^2 = p$, and hence $p =0$ or $p = 1$. In particular, this implies that $P\geq 0$. 

Such an operator $P$ projects all vectors orthogonally on the subspace $P\mathcal{H} \equiv \{P\ket{\psi}:\ket{\psi}\in\mathcal{H}\} \equiv \{\ket{\psi}\in\mathcal{H}:P\ket{\psi}=\ket{\psi}\}$.
Given any normalised state $\ket{\phi}$, we can define the rank-one operator 
\[
P_{\phi}=\ket{\phi}\bra{\phi}
\] 
as a projector on the subspace spanned by $\ket{\phi}$. It maps any vector $\ket \psi$ to $P_\phi \ket \psi = \braket \phi \psi \ket \phi$.

This notation allows us to write any self-adjoint operator $A$ (in finite dimensions) as a sum of projectors onto its eigenspaces. Let $\ket{i,k}$ be the eigenvectors of $A$ with distinct eigenvalues $a_i$, i.e., such that $A\ket{i,k}=a_i\ket{i,k}$. We define the \textit ph{spectral projectors}
\[
P_i := \sum_k \ket{i,k}\bra{i,k}
\] 
which projects on the eigenspace $P_i\mathcal{H}$ associated with the eigenvalue $a_i$. (It is easy to check that a sum of projectors is also a projector). 
This allows us to write $A$ in terms of its \textit ph{spectral decomposition}:
\[
A = \sum_i a_i P_i.
\] 
The completeness of the eigenvectors imply that $\sum_i P_i=\one$, although one of the eigenvalue $a_i$ may be equal to zero, eliminating one of the terms from the sum above. 

Observe that the probability of obtaining the value $a_i$ in a measurement of $A$ on state $\ket \psi$ can now be written simply as
\[
p_i = \sum_k |\braket{i,k}{\psi}|^2 = \bra \psi P_i \ket \psi.
\]
In particular, it depends solely on the spectral projector $P_i$ and on the state $\ket \psi$.

Note that this simple expression holds only provided that we make sure that the eigenvalues $a_i$ are distinct and hence that the projectors $P_i$ are of maximum rank. Moreover, this makes the spectral decomposition of a self-adjoint operator unique. 

Often, one is not interested in the individual outcomes probabilities when measuring an observable, but just in the average of the measured value:
\begin{mydef}
The \textbf{expectation value} $\langle A \rangle $ of an observable $A$ with spectral decomposition $A = \sum_i a_i P_i$ measured on a quantum system in state $\ket{\psi}$ is given by
\begin{equation}
\ave A = \sum_i a_i p_i = \sum_i a_i \bra{\psi} P_i \ket{\psi} = \bra{\psi} A \ket{\psi}
\end{equation}
\end{mydef}

It is worth noting that the probabilities $p_i$ themselves are expectations values of the spectral projectors:
\[
p_i = \ave{P_i}.
\]
Therefore, any prediction of quantum theory is given by the expectation value of \textit{some} self-adjoint operator.


\subsection{Generalized States}

In what follows, we introduce two settings which show that the concept of a state needs to be generalized from vectors in a Hilbert space to certain operators with special properties. We start by giving two situations motivating this generalisation, and then give an abstract definition which encompasses both examples. 

\subsubsection{Ensembles}
\label{ensembles}

Consider the case where the observer of quantum system is uncertain about what its exact state is. A natural way to model this situation is to assign probabilities $p_i$ to quantum states $\psi_i$ according to his belief about the system. This defines the \textit{ensemble} $\{(p_i,\psi_i)\}_{i=1}^n$.

The expectation value of the observable $A$ must then be the average, in terms of the classical probability distribution $i \mapsto p_i$, of the various quantum expectations values $\expectop{\psi_i}{A}$:
\begin{align}
\expect{A}&=\sum_i p_i \expectop{\psi_i}{A}.
\end{align}
This can be rewritten in a more compact form using the trace operator $\tr$. Recall that the trace of a matrix is cyclic, i.e., $\tr(AB) = \tr(BA)$. Of course, the product $AB$ must be a  square matrix, otherwise the trace is not defined. However, $A$ and $B$ themselves can be rectangular matrices. The best way of thinking about a ket $\ket{\psi_i}$ is as a matrix with just a single column, whereas a bra $\ket{\psi_i}$ is a matrix with just a single row. Considering that the trace of a number (i.e. a one-by-one matrix) is that number itself, we obtain
\begin{align}
\expectop{\psi_i}{A} = \tr \expectop{\psi_i}{A} = \tr \bigl({ \proj {\psi_i} A }\bigr).
\end{align}
We invite the reader to verify that this is correct.
This allows us to rewrite the expectation value of $A$ as
\begin{align}
\expect{A} &= \sum_i p_i \expectop{\psi_i}{A}= \sum_i p_i \tr \Bigl({ \proj {\psi_i} A }\Bigr)
= \tr(\rho A)
\end{align}
where we have defined the operator
\begin{equation}
\rho := \sum_i p_i \proj {\psi_i}.
\end{equation}
which is usually called \textit{density matrix}, or \textit{density operator}. Given that, as noted at the end of Section~\ref{positive:operators}, all quantum predictions take the form of the expectation value of \textit{some} self-adjoint operator, this means that in this scenario, the matrix $\rho$ is all that we need to know about the ensemble $\{(p_i,\psi_i)\}_{i=1}^n$ in order to compute predictions. 

For instance, considering an observable in its spectral decomposition $A=\sum_j a_j P_j$, the probability $q_j$ of observing the outcome $a_j$ is also simply
\begin{equation}
\ave{P_j} = \Tr(\rho P_j). 
\end{equation}


It is important to not that many different ensembles give rise to the same density matrix $\rho$:
\begin{theo}
Two ensembles $\{(p_i,\ket{\psi_{i}})\}_{i=1}^n$ and $\{(q_i,\ket{\phi_{i}})\}_{i=1}^n$ are represented by the same density operator, i.e,
\begin{equation}
\label{ensemble:equiv}
\sum_i p_i \proj {\psi_i} = \sum_i q_i \proj {\phi_i}
\end{equation}
if and only if there exists an unitary matrix $u_{ij}$ such that
\begin{equation}
\label{ensemble:equiv2}
\sqrt{p_i} \ket{\psi_i} = \sum_j u_{ij} \sqrt{q_j} \ket{\phi_j}.
\end{equation}
\end{theo}
\proof
The sufficiency of the condition can be verified by simple substitution and using the unitarity of the matrix $u_{ij}$, namely the fact that $\sum_i \overline u_{ik} u_{ij} = \delta_{kj}$. 
For the converse, we follow Ref.~\cite{nielsen}. Let $\rho = \sum_i p_i \proj {\psi_i}$. 
First, observe that we can always build an ensemble of mutually orthogonal states representing $\rho$ using its eigenvectors $\ket k$ and eigenvalues $p_i$: $\rho = \sum_i p_i \proj i$. If we can prove the theorem in the case where one of the ensembles is composed of orthogonal states, then we obtained the desired unitary matrix by multiplying the unitary relating the vectors of the first ensemble to the orthogonal ones with that relating the orthogonal vectors to the vectors of the second ensemble.
Let us define $\ket{a_i} := \sqrt{p_i} \ket{\psi_i}$ and $\ket{b_i} := \sqrt{q_i} \ket{\phi_i}$ for conciseness. We assume that the states $\ket a_i$ are all mutually orthogonal. In this case, observe that if $\ket \psi$ is orthogonal to all vectors $\ket a_i$, then 
\[
0 = \bra \psi \rho \ket \psi = \sum_i |\braket {b_i}{\psi}|^2
\]
which implies that $\braket {b_i}{\psi} = 0$ for all $i$, and hence $\ket \psi$ is also orthogonal to all vectors $\ket {b_i}$. That statement also holds if we exchange the two families of vectors. This means that the families $\ket {a_i}$ and $\ket {b_i}$ have the same span. Hence there are complex numbers $c_{ij}$ such that
\[
\ket{b_i} = \sum_j c_{ij} \ket {a_j}.
\]
Moreover, we have
\[
\rho = \sum_k \ketbra{a_k}{a_k} = \sum_k \ketbra{b_k}{b_k} = \sum_{ij} \sum_k c_{ki} \overline {c_{kj}} \ketbra{a_i}{a_j}.
\]
From the linear independence of the matrices $\ketbra{a_i}{a_j}$, we conclude that $\sum_k c_{ki} \overline {c_{kj}} = \delta_{ij}$, and hence the matrix $c_{ij}$ is unitary.
\qed

If the two ensembles have a different number of vectors, one can simply pad the smaller ensemble with vectors corresponsing to zero probabilities. 


The uncertainty involved in a given ensemble can be measured by the \textit{Shannon entropy} $S(p) = - \sum_i p_i \ln p_i$. For a given density matrix $\rho$, this entropy is not defined because it depends on the ensemble from which $\rho$ is constructed. However, it makes sense to consider the ensemble that corresponds to a minimal uncertainty. This defines the \textit{von Neumann entropy} associated with a density matrix:
\begin{equation}
S(\rho)=-\Tr(\rho\ln \rho)=\min_{\rho = \sum_i p_i \proj {\psi_i}} S(p),
\end{equation}
where the minimum is over all possible ensembles $\{(p_i,\psi_i)\}_{i=1}^n$ such that $\rho = \sum_i p_i \proj {\psi_i}$. In fact, one can show that this minimum is reached whenever the states $\psi_i$ are all orthogonal. In this case, the probabilities $p_i$ are simply the eigenvalues of $\rho$. Therefore, $S(\rho)$ is the Shannon entropy computed from the eigenvalues of $\rho$. We will come back to this special diagonalising ensemble in Section~\ref{density:matrix}.


\subsubsection{Tensor Product and Reduced States}
\label{reduced:state}


For this example, we need to introduce another tools in quantum theory which allows us to compose several systems into a bigger one, or, alternatively, to consider subsystems of a larger system. 

Consider two quantum systems $A$ and $B$ with respective Hilbert spaces $\hilbert_A$ and $\hilbert_B$. 
Let $\{\ket{i}_A\}_{i=1}^n$ be an orthonormal basis of $\hilbert_A$ and $\{\ket{j}_B\}_{i=1}^m$ an orthonormal basis of $\hilbert_B$ respectively. We define the larger Hilbert space
\begin{equation}
\hilbert = \hilbert_A \tensor \hilbert_B
\end{equation}
of dimension $n \cdot m$, with orthonormal basis $\ket{i,j}$, $i=1,\dots,n$, $j=1,\dots,m$. It is convenient to write formally
\begin{equation}
\ket{i,j} \equiv \ket{i}_A\tensor\ket{j}_B.
\end{equation}
Hence, any vector $\psi \in \hilbert$ can be expanded as 
\begin{equation}
\ket{\Psi}=\sum_{i=1}^n \sum_{j=1}^m\Psi_{ij}\ket{i}_A\tensor\ket{j}_B.
\end{equation}

We can think of $\hilbert_A$ and $\hilbert_B$ as two ``independent'' parts of $\hilbert$. In particular, given any vectors $\psi_A \in \hilbert_A$ and $\psi_B  \in \hilbert_B$, we can construct the joint vector
\begin{equation}
\ket{\psi_A} \otimes \ket{\psi_B} = \sum_{i=1}^n \sum_{j=1}^m \braket i {\psi_A} \braket j {\psi_N}  \, { \ket i}_A \otimes {\ket j}_B.
\end{equation}

Moreover, there is a natural way to map any operator $A$ acting on $\hilbert_A$, or $B$ acting on $\hilbert_B$, to an operator on $\hilbert$, respectively written as $A \otimes \one$ and $\one \otimes B$. They are defined by their action on the basis as follows:
\begin{align}
(A \otimes \one) (\ket i_A \otimes \ket j_B) &= (A {\ket i}_A) \otimes {\ket j}_B\\
(\one \otimes B) (\ket i_A \otimes \ket j_B) &= {\ket i}_B \otimes (B {\ket j}_B).
\end{align}
Observe that the labels $A$ and $B$ on the kets are not really necessary, as we always take care to preserve the ordering of the tensor factors: system $A$ on the left of the tensor product and system $B$ on the right. Hence in the following we would write the above two equations as
\begin{align}
(A \otimes \one) (\ket i \otimes \ket j) &= (A {\ket i}) \otimes {\ket j}\\
(\one \otimes B) (\ket i \otimes \ket j) &= {\ket i} \otimes (B {\ket j}).
\end{align}

An essential property of this representation of the operators $A$ and $B$ is that they commute:
\begin{equation}
(A \otimes \one)(\one \otimes B) = (\one \otimes B) (A \otimes \one) = A \otimes B.
\end{equation}
We also have the following algebraic properties,
\begin{align}
(A \otimes B)(\ket \psi \otimes \ket \phi) &= (A\ket \psi) \otimes (B\ket \phi)\\
(A \otimes B)(A' \otimes B') &= (AA') \otimes (BB')\\
\alpha(A \otimes B) &= (\alpha A)\otimes B = A \otimes (\alpha B)\\
(A \otimes B)^\dagger &= A^\dagger \otimes B^\dagger,
\end{align}
for any operators $A,A',B,B'$ acting on the respective Hilbert spaces, and any $\alpha \in \mathbb C$. 

The tensor product can be straightforwardly generalised to more than two tensor factors, and two non-square matrices, i.e., operators between two different Hilbert spaces. In particular, the tensor product of two kets is the same as that of two matrices with only one column (columns vectors). We invite the reader to experiment with this concept. In particular, a good exercise is to understand the tensor product in terms of its action on matrix components. Here we just observe that the tensor product is associative:
\begin{align}
(A \otimes B) \otimes C \equiv A \otimes ( B \otimes C ) \equiv A \otimes  B \otimes C.
\end{align}

Also, a construction that we will often encounter is the tensor product of an operator and ket, or a bra, whose action on states is defined as follows:
\begin{align}
( A \otimes \ket{\psi} )  \ket{i} &= A \ket{i}  \otimes \ket{\psi} \\
( A \otimes \bra{\psi} )  ( \ket{i} \otimes \ket{j}) &= \braket{\psi}{j} A \ket{i}.
\end{align}

We are now in measure to define the concept of \textit{reduced state}.
Suppose that, for some reason, we decide to only measure observables on $\hilbert_A$, i.e., observables represented by operators of the form  $A\tensor\one$, where $A$ is any operator on $\hilbert_A$. Observe that, in particular, if the spectral decomposition of $A$ is $A = \sum_i a_i P_i$, then the spectral decomposition of $A\tensor\one$ is $A = \sum_i a_i P_i\tensor\one$. Indeed, it is easy to check that $P_i \otimes \one$ are also projectors. 

This implies that all the possible predictions are of the form of the expectation value of a self-adjoint operator having the shape $X \otimes \one$, which, on an arbitrary state $\Psi \in \hilbert_A \otimes \hilbert_B $, is
\begin{equation}
\av{X\tensor\identity}=\expectop{\psi}{X\tensor\identity}=\Tr\big((X\tensor\identity)\ket{\psi}\bra{\psi}\big)=\sum_j \Tr\Big(\bigl({X\tensor\ket{j}\bra{j}}\bigr)\ket{\psi}\bra{\psi}\Big),
\label{eq:partialob}
\end{equation}
where, in the last step, we expanded the identity on system $B$ in terms of a basis with elements $\ket j$. Observing that 
\begin{equation}
X \otimes \proj j = (\one \otimes \ket j) X (\one \otimes \bra j),
\end{equation}
and using the cyclicity of the trace, we obtain
\begin{equation}
\begin{split}
\av{X\tensor\identity} 
&=\sum_j \Tr\Bigl({ (\one \otimes \ket j) X (\one \otimes \bra j)\ket{\psi}\bra{\psi}}\Bigr)\\
&= \sum_j \Tr\Bigl({ X (\one \otimes \bra j)\ket{\psi}\bra{\psi} (\one \otimes \ket j)  }\Bigr)\\
&= \Tr( X \rho)
\end{split}
\end{equation}
where we defined the operator
\begin{equation}
\rho = \sum_j (\one \otimes \bra j)\ket{\psi}\bra{\psi} (\one \otimes \ket j).\\
\end{equation}
This operator $\rho$, which is also a \textit{density matrix}, is an operator acting only on Hilbert space $\hilbert_A$. Nonetheless, it contains all the information that we will ever need about the state $\psi \in \hilbert_A \otimes \hilbert_B$, provided we are restricted to only measure observables of system $A$.

It will be useful to define more generally the operation which maps $\proj \psi$ to $\rho$. It is a linear map that we call the \textit{partial trace over system $B$}, written $\tr_B$, which acts on operators as follows
\begin{equation}
\tr_B(Z) = \sum_j (\one \otimes \bra j) Z (\one \otimes \ket j).
\end{equation}
It takes an operator $Z$ acting on $\hilbert_A \otimes \hilbert_B$ to an operator acting on $\hilbert_A$ alone. 

This is called partial trace because the full trace $\tr$ over $\hilbert_A \otimes \hilbert_B$  is given by
\begin{equation}
\tr = \tr_A \tr_B = \tr_B \tr_A, 
\end{equation}
where the implicit product used here is simply the composition of maps.

The above reasoning can be carried likewise if we started with a density matrix $\rho_{AB}$ on system $\hilbert_A \otimes \hilbert_B$  rather than simply the vector $\psi$:
\begin{mydef}
Let $\rho_{AB}$ be a density matrix on a bipartite system, then 
\begin{equation}
\rho_A = \tr_B(\rho_{AB})  
\label{eq:reduced_state}
\end{equation}
is the \textit ph{reduced state} on system $A$.
\end{mydef}

\subsubsection{Density Operator/Matrix}
\label{density:matrix}

In the previous two sections, we have seen that it is useful to characterize the state of a system through an operator $\rho$, such that the expectation value of a self-adjoint operator $A$ can be obtained via $\Tr(\rho A)$. Now we take this more general concept of states to its full generality:
\begin{mydef}
A density operator (\textbf{state}) is an operator $\rho$ satisfying
\begin{enumerate}[(i)]
\item $\rho\geq 0$, 
\item $\Tr{\rho}=1$.
\end{enumerate}
The expectation value of the observable represented by the self-adjoint operator $A$ with respect to the state $\rho$ is given by
\begin{equation}
\langle A\rangle = \Tr(\rho A).
\end{equation}
\end{mydef}

These two conditions completely capture the concept of a density matrix that emerged in either of the two contexts studied in Section~\ref{ensembles} or \ref{reduced:state}. That is, not only do the density matrices emerging in these context always satisfy conditions (i) and (ii), but also any matrices satisfying these condition can emerge in both contexts.

First, observe that these conditions are necessary and sufficient for the following interpretation: for any observable represented by a self-adjoint operator $A$ with spectral decomposition $A = \sum_i a_i P_i$, we want the numbers $p_i = \tr(\rho P_i)$ to form a probability distribution, i.e., that $p_i > 0$ and $\sum_i p_i = 1$. Indeed, this requirement implies in particular that for any vector $\ket \psi$ we must have $\tr(\rho \proj \psi) = \bra \psi \rho \ket \psi \ge 0$, which simply means $\rho \ge 0$, namely Condition (i). Moreover, from the fact that $\sum_i P_i = \one$,  $\sum_i p_i = 1$ directly implies $\tr (\rho \one) = 1$, which is Condition (ii). The converse is similarly straightforward. 


Moreover, any such density matrix can be obtained either as an ensemble or as a reduced state, which shows that this definition is not taking us away from the accepted framework of quantum mechanics. 
Indeed, suppose $\rho$ is any matrix satisfying conditions (i) and (ii). Let $\ket i$ be a complete set of eigenvectors for $\rho$ with eigenvalues $p_i$, $i=1,\dots,n$. The two conditions imply that $p_i \ge 0$ and $\sum_i p_i = 1$, therefore $\rho$ represents the ensemble $\{p_i,\ket{i}\}_{i=1}^n$ via its spectral decomposition $\rho=\sum_i p_i \proj{i}$. 
This arbitrary density matrix can also represent a reduced state. Indeed, consider two copies of the Hilbert space on which it is defined, and the bipartite vector $\ket \psi = \sum_i \sqrt{p_i} \ket i \otimes \ket i$. One can check that $\rho$ is obtained by tracing out $\proj \psi$ on any of the two systems. The vector $\ket \psi$ is generally called a \textit{purification} of $\rho$. 



\begin{mydef}
We call a state $\rho$ \textbf{pure} if it cannot be written as an ensemble of two or more distinct states.
(i.e., it is extremal in the convex set of density matrices). The following propositions are all equivalent (on finite-dimensional Hilbert spaces):
\begin{enumerate}[(i)]
\item $\rho$ is pure
\item $S(\rho) = 0$
\item $\rho^2=\rho$
\item $\rho = \proj \psi$ for some normalised vector $\psi$.
\end{enumerate} 
\end{mydef}
\proof We show $(i) \Rightarrow (ii) \Rightarrow (iii) \Rightarrow (iv) \Rightarrow (i)$. It directly follows from the extremality condition that $S(\rho) = 0$. If $\rho = \sum_i p_i \proj i$ for orthogonal eigenstates $\ket i$, then we have $S(\rho) = - \sum_i p_i \ln p_i = 0$. But each term $-p_i \ln p_i$ is positive since $0 \le p_i \le 1$. Hence, for all $i$ we have $- p_i \ln p_i = 0$, which is true either if $p_i = 0$ or $\ln p_i = 0$, i.e., $p_i = 1$. Hence $\rho$ is a projector: $\rho^2 = \rho$. Moreover, since $\sum_i p_i = 1$ then only a single $p_i$ can be nonzero. It follows that $\rho = \proj i$. Moreover, such a state is extremal because $\proj \psi = p \rho_1 + (1-p) \rho_2$ implies, by multiplying both sides by $\bra \psi$ on the left and $\ket \psi$ on the right, that $1 = p \bra \psi \rho_1 \ket \psi + (1-p) \bra \psi \rho_2 \ket \psi$, or $p (1 - \bra \psi \rho_1 \ket \psi) + (1-p) (1 - \bra \psi \rho_2 \ket \psi) = 0$. Since both terms are positive we must have, $\bra \psi \rho_i \ket \psi = 1$, which implies that $\rho_1 = \rho_2 = \proj \psi$. \qed


We close this section with some examples:
\begin{ex}[{Entanglement}] \label{ex:entangled}
Consider a bipartite system, with Hilbert space $\hilbert_A\otimes\hilbert_B$. Let $\{\ket{i}_A\}_{i=1}^n$ and $\{\ket{j}_B\}_{j=1}^n$ denote orthonormal bases of $\hilbert_A$ and $\hilbert_B$ respectively. The pure state $\ket{\Omega}=\frac{1}{\sqrt{n}}\sum_i\ket{i}_A\otimes\ket{i}_B$ corresponds to the density matrix 
\[
\rho_{AB}=\ket{\Omega}\bra{\Omega}=\frac{1}{n}\sum_{i,j}\left(\ket{i}_A\otimes\ket{i}_B\right)\left(\bra{j}_A\otimes\bra{j}_B\right) = \frac{1}{n}\sum_{i,j} {\ketbra i j}_A\otimes {\ketbra i j}_B. 
\]
The reduced state on system $A$ is obtained by the partial trace:
\begin{align}
\rho_A&=\Tr_B(\rho_{AB})=\sum_k\bigl(I\tensor\bra{k}\bigr)\rho_{AB}\bigl(I\tensor\ket{k}\bigr)\\
      &=\frac{1}{n}\sum_{k,i,j}\bigl(I\tensor\bra{k}\bigr)\bigl({\ketbra i j}\otimes{\ketbra i j}\bigr)\bigl(I\tensor\ket{k}\bigr)\\
			&=\frac{1}{n}\sum_{k,i,j}I\ket{i}\bra{j} I\underbrace{\braket{k}{i}}_{\delta_{k,i}}\underbrace{\braket{j}{k}}_{\delta_{j,k}} =\frac{1}{n}\sum_{k}\ket{k}\bra{k}=\frac{\one}{n}.
\end{align}
The reduced state $\rho_A = \one/n$ is referred to as the \textit ph{maximally mixed} state. Because it is invariant under any unitary transformation $U$, it is also the reduced state of the states of the form $(U \otimes \one)\ket \Omega$, which just amount to a different choice of basis on system $A$. 
Observe that, although the full state $\rho_{AB} = \proj \Omega$ is pure, and hence $S(\proj \Omega) = 0$, i.e., we possess maximal information about it, its ``part'' $\rho_A$ has maximal entropy: $S(\rho_A) = \ln n$, 
which means that we know absolutely nothing about the state of system $A$. This is completely contrary to classical systems, where knowing the state of the whole system implies also complete knowledge of its parts. Quantum states having this non-classical property are called entangled. Hence, we call any \textit ph{pure state} $\rho_{AB}=\proj{\psi}$ of the compound system \textit ph{entangled} whenever its parts have non-zero entropy. Since the parts of $\ket \Omega$ have maximal entropy, $\ket \Omega$ is maximally entangled. 
\end{ex}
 
\begin{ex}[{Bloch Sphere}]
We characterise the complete set of states one the two dimensional Hilbert space $\hilbert=\mathbb{C}^2$. For this purpose it is useful to introduce the Pauli matrices $\{\sigma_0,\sigma_1,\sigma_2,\sigma_3\}$, which form a basis of the complex \textit ph{vector space} of two-by-two complex matrices:
\begin{equation}
\sigma_0=\left(
           \begin{array}{cc}
             1 & 0 \\
             0 & 1 \\
           \end{array}
         \right),\quad
\sigma_1=\left(
           \begin{array}{cc}
             0 & 1 \\
             1 & 0 \\
           \end{array}
         \right),\quad
\sigma_2=\left(
           \begin{array}{cc}
             0 & -i \\
             i & 0 \\
           \end{array}
         \right),\quad
\sigma_3=\left(
           \begin{array}{cc}
             1 & 0 \\
             0 & -1 \\
           \end{array}
         \right).
\end{equation}

The set of Hermitian matrices with trace one is completely parametrised with three real numbers $r_1, r_2, r_3$ as:
\begin{equation}
\rho=\frac{1}{2}\sigma_0+\frac{1}{2}(r_1\sigma_1+r_2\sigma_2+r_3\sigma_3)=
\frac{1}{2}
\left(
  \begin{array}{cc}
    1+r_3 & r_1-ir_2 \\
    r_1+ir_2 & 1-r_3 \\
  \end{array}
\right),\quad
r_1,r_2,r_3 \in \mathbb{R}.
\end{equation}
Moreover, for the operator $\rho$ to be positive, its eigenvalues $\lambda_1,\lambda_2$ need to be non-negative. Since we already guaranteed that they sum to one, they cannot be both negative. Therefore, the only extra condition required is that their product be non-negative, that is, 
\begin{equation}
\det{\rho}=\lambda_1 \lambda_2=\frac{1}{4}(1-r_3^2-(r_1^2-(ir_2)^2))=\frac{1}{4}(1-r_1^2-r_2^2-r_3^2)=\frac{1}{4}(1-||\vec{r}||^2) \ge 0.
\end{equation}
 From this we can see that the set of possible states (density matrices) corresponds via the above parametrisation to the 3-dimensional ball characterised by $\|\vec r\|^2 \le 1$. This is manifestly a convex set whose extreme points; the pure states, lie on its boundary: the unit sphere. 
\end{ex}

%% file: Observables.tex
\subsection{Generalised propositions}

In the previous section we generalized the notion of state. We will also need to generalize the concept of observable. For this purpose, we first consider the most elementary notion of \emph{yes/no} observables, i.e., observables with only two distinct outcomes. Following the tradition of formal logic, one may conceive it as a \emph{proposition} about the physical system; a formal statement which can be either true or false. 

According to the standard formulation of quantum theory, such an observable is represented by a self-adjoint operator $A$ with two distinct eigenvalues $a \neq b$. Its spectral decomposition has the form:
\begin{align}
A&=aP+bQ 
\end{align}
where $P$ and $Q$ are two orthogonal projectors such that $P + Q = \one$, i.e., $P = \one - Q$. 
The particular values of $a$ and $b$ do not matter as long as they are distinct. In particular, we could have chosen $a = 1$ and $b = 0$, so that we simply have $A \equiv P$. We can then interpret that, in the outcome of a measurement of $P$, obtaining the eigenvalue $1$ signifies that the proposition $P$ is \emph{true}, and obtaining the eigenvalue $0$ signifies that it is \text{false}. The converse proposition, its negation, is then simply characterised by the observable $P^\perp = \one - P$. 
A state $\rho$ of the system assign probabilities to the truth values of these propositions:
\begin{align}
\text{prob(P is true)}&=\Tr(\rho P) \label{eq:probabilities1} \\ 
\text{prob(P is false)}&=\Tr(\rho P^{\perp})=\Tr(\rho(\one-P))=1-\Tr(\rho P)
 \label{eq:probabilities2}
\end{align}

Like in the previous chapter, where we introduced uncertainty about our knowledge about the state, one can also add uncertainty to our knowledge of the measurement device. One could imagine that, with probability $p$ our measurement device shows the wrong measurement outcome. This means that we must review our probability \eqref{eq:probabilities1} as follows:
\begin{align}
\text{prob(P is true)} &=  (1-p) \Tr(\rho P) + p \Tr(\rho P^\perp) \\
 &=\Tr\bigl({ \rho ( (1-p) P + p P^\perp  )}\bigr) \\
 &=\Tr(\rho E),
\end{align}
where we have defined the operator
\begin{equation}
E := (1-p) P + p P^\perp
\end{equation}
Moreover, \eqref{eq:probabilities2} changes accordingly to
\begin{align}
\text{prob(P is false)}&= p \Tr(\rho P) + (1-p) \Tr(\rho(\one -P)) \\
 &=\Tr(\rho(\one -E)).
\end{align}
We see that the operator $E$ is not a projector in general, but it is still positive $E\geq 0$, and $\Tr(\rho E)$ directly yields the \emph{probability} of the outcome of a measurement. We can therefore generalize the concept of \emph{proposition}, allowing them to be represented by any operators satisfying $0\leq E \leq \one $, rather than simply projectors. 

These operators are traditionally called \emph{effects}. It is important to observe that taking the spectral decomposition of $E$ would have no physical meaning unlike for an observable. 

Also we note that projectors are a special type of effect, which we call \emph{sharp}, because they do not introduce extra uncertainty beyond that represented by the state. 


 
\subsection{Abstract State/Effect Formalism}

Before we discuss the properties of quantum effects in more details, we introduce the abstract state/effect formalism which will set the language for a framework which unifies classical and quantum theories. 
It is known as the \emph{state/effect} formalism, or generalised probability theory. 

Such a theory consists of a set of \emph{effects} (propositions about the systems which can be true or false) and a set of \emph{states} (an assignment of a probability to each effect). This is essentially a Bayesian point of view on physics, where a \emph{state} represents an observer's state of knowledge about a system. It may or may not represent an objective attribute of a physical system, depending on whether the observer is actually right, or whether such objective attributes even exist. 

	
The most basic mathematical structure that one usually requires these sets of effects and observables to possess is one that gives the ability to take convex combinations of objects. E.g., if $\rho_1$ and $\rho_2$ are two states, then one can form a new valid state as $p \rho_1 + (1-p) \rho_2$ where $p \in [0,1]$. This corresponds to the physical interpretations of a mixture: the observer is uncertain whether the state is $\rho_1$ or $\rho_2$, and attributes probability $p$ to it being $\rho_1$. The same can be done with effects as in the previous section. 


The following definition characterizes special points of the convex set of states:
\begin{mydef}
Let $\Omega$ be the set of states. Then the element $\rho\!\in\! \Omega$ is \textbf{extremal}, or \textbf{pure}, if does not admit a proper convex combination, i.e., if there exists $0 < p < 1$ such that
\begin{equation}
\rho = p  \rho_1 + (1-p)\rho_2,
\end{equation}
then one must have $\rho_1 = \rho_2 = \rho$.
\end{mydef}


As seen in the previous section, quantum theory in finite dimension can be formulated in this language as follows:
\begin{ex}[{Quantum Theory}]
Given a finite-dimensional Hilbert space $\hilbert$, we identifies states and effects with certain operators as follows:
\begin{align*}
\emph{states}\ \rho &: \ \rho \geq 0 \ , \ \Tr(\rho)=1 \\ 
\emph{effects}\ E &: \ 0 \leq E \leq \one , 
\end{align*}
where a state $\rho$ assigns to the effect $E$ the probability: $p(\text{$E$ is true}) =\Tr(\rho E)$. 

This defines the complete structure of finite-dimensional quantum theory, which is essentially the Born rule. We will see below how the other standard postulates, such as ``state collapse'' can be derived from this structure. 

One can check that the extreme points of the set of states are of the form $\rho = \proj \psi$, $\psi \in \hilbert$, and the extreme points of the set of effects are the projectors: $E^2 = E$.
\end{ex}
 
Another example that fits into this framework is classical probability theory:
\begin{ex}[{Classical Probability Theory}]
Let $\Omega$ be a finite set. We define states and effects as
\begin{align*}
\emph{states}&: \ \rho=\{p_i\}_{i\in\Omega} \ , 0\leq p_i\leq 1 \ , \ \sum_i p_i=1  \\ 
\emph{effects}&: E=\{e_i\}_{i\in\Omega} \ , \ 0\leq e_i\leq 1 ,
\end{align*}
where a state $\rho$ assigns to the effect $E$ the probability: $p(\mbox{E is true})=\sum_i p_i e_i$. 

The states are characterized by a probability distribution on the set $\Omega$. Extremal states are given by $p_i=\delta_{ij}$, and are hence one-to-one with elements of $\Omega$. 
 
Normally, propositions are characterised as subsets $\omega \subseteq \Omega$: namely those pure states for which the proposition is true. These corresponds to the effects associating $1$ with elements of $\omega$ and $0$ with elements not in $\omega$. They are the extremal points of the full set of effects. 

For example let $\Omega=\mathbb{Z}$. If this system characterises a ``random variable'' called $X$, then the proposition ``$X$ is positive'' corresponds to the effect $E = \left\{e_i\right\}_{i\in\mathbb{Z}}$ with
\begin{equation}
e_i = \begin{cases} 
1 & \text{if $i\geq 0$},\\
0 & \text{otherwise},
\end{cases}
\end{equation}
or, more generally, the proposition ``$X$ belong to some subset $\omega\subset\Omega$'' corresponds to the effect
\begin{equation}
e_i= \begin{cases} 
1 & \text{if $i\in \omega$},\\
0 & \text{otherwise}.
\end{cases}
\end{equation}
Evaluating the probability for the latter amounts to 
\begin{equation*}
p(\mbox{E is true})=p(X\in\omega)=\sum_i e_i p_i=\sum_{i\in\omega}p_i,
\end{equation*}
as expected from standard probability theory.
\end{ex}

This framework allows us to formulate classical probability theory in terms of a quantum theory with additional restrictions: Let $\{p_i\}_{i\in\Omega}$ denote a classical state and $E=\{e_i\}_{i\in\Omega}$ a classical effect on an outcome set $\Omega$. Enforcing that the states $\rho$ and Effects $E$ are diagonal in a fixed basis $\{\ket{i}\}_{i=1}^n$ allows for the  following identification:
\begin{align*}
\rho&=\sum_i p_i \proj{i}\\
E&=\sum_i e_i \proj{i}.
\end{align*}
Therefore, this amounts to valid state/effect configuration of a quantum system, where the probability assigned to the effect is exactly the same as in the classical setting $p(\mbox{E is true})=\Tr(\rho E)=\sum_i p_i e_i$.


\subsection{Algebraic Formulation}

The state/effect formalism allows us to introduce a formalism which generalises both quantum and classical theories, while keeping as much structure as possible from both. 
Moreover, this framework allows for infinite-dimensional generalizations even beyond Hilbert spaces, though in this lecture we focus mostly on finite systems for pedagogical reasons. 

For this purpose, recall the notion of an \emph{algebra} $\mathscr{A}$ over $\mathbb{C}$. It is a  complex vector space on which a multiplication is defined. This multiplication is compatible with addition, i.e.\ $a(b+c)=ab+ac$ for $a,b,c\in\mathscr{A}$. The multiplication is called \emph{associative} if also $a(bc)=a(bc)$. Also, it is called \emph{unital} if it contains an identity $\one $ for the multiplication: $\one a=a\one =a$. Based on these structures we define
\begin{mydef}
A \textbf{$*$-algebra} $\mathscr{A}$ is an associative algebra with an anti-linear map $*:\mathscr{A}\rightarrow\mathscr{A}$ satisfying the following properties $\forall a,b\in\mathscr{A}$:
\begin{itemize}
	\item $\left(a^*\right)^*=a$
	\item $\left(ab\right)^*=b^*a^*$
\end{itemize}
Moreover it is \textbf{unital} if it contains an \emph{identity} $\one$ such that, $\forall a \in\mathscr{A}$,
	$\one a = a \one = a$.
\end{mydef}
A prominent example of a $*$-algebra is given by the complex matrices where the involution~$*$ is given by the conjugate-transpose operation~$\dagger$. Of course, there is a natural concept of substructure of $*$-algebras:
\begin{mydef}
Let $\mathscr{B}$ be a $*$-algebra and $\mathscr{A}\subset\mathscr{B}$. $\mathscr{A}$ is a \textbf{$*$-subalgebra}, if the following properties hold:
\begin{itemize}
	\item $\mathscr{A}$ is a linear subspace: $a,b\in\mathscr{A}, \ \alpha,\beta\in \mathbb{C} \ \Rightarrow \ \alpha a +\beta b \in \mathscr{A}$
	\item $\mathscr{A}$ is closed under multiplication: $a,b\in\mathscr{A} \ \Rightarrow  \ ab\in\mathscr{A}$
	\item $\mathscr{A}$ is closed under conjugation: $a\in\mathscr{A}\ \Rightarrow  \ a^*\in\mathscr{A}$
\end{itemize}
\end{mydef}

The structure of a $*$-algebra allows us to introduce a concept of positivity that generalises that of operators:
\begin{mydef} 
\label{def:positive} 
Let $\mathscr{A}$ be a $*$-algebra. 
 We say $a\! \in\! \mathscr{A}$ is \textbf{positive}, i.e. $a\geq0$, if there are elements $b_i \in \mathscr A$, $i=1,\dots,n$ such that
\[
a = \sum_i b_i^* b_i.
\]
\end{mydef}
This definition is such that the set of positive elements is a \emph{cone}, namely closed under linear combinations with positive scalars. In turn, this makes the relation $a \le b$ defined by $b -a \ge 0$ into a partial order.

It is not hard to see that for matrices, with $A^* := A^\dagger$, this condition is equivalent to Definition~\ref{positive:operators}. For instance, if $A$ is positive as a matrix, we have seen that $A = \sum_i a_i \proj i$, where $a_i>0$. Hence, using $B_i := \sqrt{a_i} \bra i$ we have $A = \sum_i B_i^\dagger B_i$.

Using this structure, we can associate a generalized probability theory to any unital $*$-algebra $\mathscr{A}$:
\begin{mydef}
Let $\algebra$ be a (possibly infinite-dimensional) unital $*$-algebra. We define the \textbf{effects} associated with $\mathscr A$ to be the positive elements which are smaller or equal to the identity
\begin{align*}
\text{effects: } e\in\algebra,  \ 0 \leq e \leq \one 
\end{align*}
Moreover, we define the \textbf{states} as certain linear functionals on $\algebra$:
\begin{align*}
\text{states: } f: \mathscr{A}\rightarrow \mathbb{C},\, \text{ $f$ linear, positive and unital},
\end{align*}
where positivity of $f$ means that $f(a)\ge 0$ for all $a\ge 0$, and unitality is the normalisation condition $f(\one) = 1$.
\end{mydef} 
Note, that this assignment gives a physical meaning to the abstract mathematical object of a $*$-algebra, once some effects are associated with actual experimental setups.

Before we describe examples, we need the following correspondence between linear functionals on some finite dimensional algebra and elements inside the algebra.
\begin{theo} 
Suppose that $\algebra$ is a $*$-subalgebra of the $n$-by-$n$ complex matrices $\mathscr M_n$: $\algebra\subset M_n$. Then for every linear functional $f:\algebra \rightarrow \mathbb{C}$ there exists $R\in M_n(\mathbb{C})$ such that
\begin{equation}
f(A)=\Tr(RA) \quad \forall A\in \algebra
\end{equation}
\label{th:riesz}
\end{theo}
\proof
Since $\one = \sum_i \proj i$, we have $A = \sum_{ij} \bra i A \ket j \ketbra i j$. Therefore, $f(A)=\sum_{i,j}\bra i A \ket j f(\ket{i}\bra{j})$. If we define $R := \sum_{ij} f(\ket{i}\bra{j}) \ketbra j i$, then $f(\ket{i}\bra{j}) = \bra j R \ket i$, and
$
f(A) = \sum_{i,j}  \bra i A \ket j \bra j R \ket i   = \tr(R A ).
$
\qed 


The previously defined probability theories can in fact be defined in this way from $*$-algebras:
\begin{ex}[{Quantum Theory}]
Given a Hilbert space $\hilbert = \mathbb C^d$, consider the unital $*$-algebra of linear maps on $\hilbert$: $\mathscr A = \mathcal B(\hilbert)$ ($\mathcal B$ stands for ``bounded linear map'', which is true of any map in finite-dimension). In this case, the star operation is given by the adjoint:
\[
A^* \equiv A^\dagger. 
\]
 This algebra is the same as that of $d$-by-$d$ complex matrices: $\mathscr{A} \equiv \mathscr M_d$.
Building the states and effects from $\mathscr A$ yields quantum theory with Hilbert space $\hilbert$. 

Indeed, the set of effects $E$ is precisely the one we previously identified in quantum theory. Therefore, we just need to show that the states defined as positive linear functionals $f:\mathcal{B}(\hilbert)\rightarrow\mathbb{C}$ correspond to density matrices. Theorem \ref{th:riesz} tells us that for every linear functional $f$ we can find a $R \in \mathscr M_d$ such that $f(A)=\Tr(RA) \ \forall A\in\mathcal{B}(\hilbert)$. The positivity of the functional $f$ says that for all positive matrices $A \ge 0$, $\Tr(RA)\geq 0$.
Choosing in particular $A=\proj{\psi}$, this says that $\bra \psi R \ket \psi \ge 0$ for all $\psi \in \hilbert$, i.e., $R$ is a positive matrix. The trace-condition $\Tr R=1$ also follows directly from $f(\one )=1$. Therefore $R$ is a density matrix. 
\end{ex}

Likewise, we also obtain classical probability theory if we consider a commutative $*$-algebra:

\begin{ex}[{Classical Probability Theory}] Given a finite set $\Omega$, we consider the unital $*$-algebra 
\[
\mathscr A = \funs(\Omega)
\]
of functions $a: \Omega \rightarrow \mathbb C$, with all algebra operations defined pairwise: for any $a,b \in \mathscr A$, $x \in \Omega$, $\alpha\, \beta \in \mathbb C$, 
\begin{align}
(\alpha a + \beta b)(x) &:=  \alpha a(x) + \beta b(x)\\
(ab)(x) &:= a(x) b(x)\\
a^*(x) &:= \overline{a(x)} 
\end{align}
where $\overline \alpha$ denotes the complex conjugate of $\alpha \in \mathbb C$.


It is easy to check that positive elements of $\mathscr A$ are those for which $a(x) \ge 0$ for all $x$, and hence, the effects are precisely those of classical probability theory. A natural linear basis of $\mathscr A$ is given by the index functions $\chi_x$ defined by $\chi_x(y) = \delta_{x,y}$ for all $x, y \in \Omega$. Hence, given a linear functional $f: \mathscr A \rightarrow \mathbb C$, we have $f(a) = \sum_x a(x) f(\chi_x)$. Let $p \in \mathscr A$ defined by $p(x) = f(\chi_x)$, we have
\(
f(a) = \sum_x a(x) p(x).
\)
If $f$ is a state, then we must $p(x) \ge 0$ for all $x$ and $f(\one) = \sum_x p(x) = 1$. Hence states are probability distributions over $\Omega$.

Observe that this algebra is equivalent to that of diagonal $d$-by-$d$ matrices, where $d$ is the number of elements in $\Omega$. Indeed, if we identify any $a \in \mathscr A$ with the diagonal matrix $A$ with diagonal elements $a(x)$, $x \in \Omega$, which we can also write using the bracket notation $A = \sum_{x \in \Omega} a(x) \proj x$, where the vectors $\ket x$, form an orthogonal basis, we see that the matrix product and other operations are the ones above. 

We will also need to consider classical systems with continuously many pure states. This can be formalised in several ways. Here, we will represent this situation by a \emph{measurable} set $\Omega$, i.e., a set with a concept of volume, or measure, for most subsets. A typical example will be $\Omega = \mathbb R$. The algebra that we will consider, 
\[
L(\Omega) \equiv L^\infty(\Omega),
\]
is the commutative algebra of \emph{bounded} functions $f: \Omega \rightarrow \mathbb C$, i.e., there is a constant $C$ such that $|f(x)| < C$ for all $x \in \Omega$ except possibly for a subset of measure zero. $L^\infty(\Omega)$ is the natural classical counterpart of $\mathcal B(\hilbert)$. 


\end{ex}

In addition to the $*$-algebra structure described above, one normally also requires a \emph{norm} $\|\cdot \|$ to be defined (and finite) on every element of the algebra, and which respects compatibility conditions with respect to the product and $*$ operations. 

If also $\|x y\| \le \|x\| \|y\|$ for all $x,y \in \mathcal A$, then $\mathcal A$ is a \emph{Banach algebra}. If, moreover, $\|x^* x\| = \|x\|\|x^*\|$ for all $x \in \mathcal A$, then $\mathcal A$ is a $C^*$-algebra, which is the most popular algebraic framework for quantum theory. We define here this important concept for the reader's benefit, but we will continue working with concrete operators on Hilbert spaces for simplicity in this document.

\begin{ex}[{$C^*$-Algebra}] 
A $C^*$-algebra is a $*$-algebra that is also a Banach algebra, i.e., complete with respect to a norm satisfying $\|x y\| \le \|x\| \|y\|$ for all $x,y \in \mathcal A$, and also such that 
\begin{equation}
\|xx^*\| = \|x\|\|x^*\| = \|x\|^2 \quad\quad \text{for all $x \in \mathcal A$.}
\end{equation}
\end{ex}

A typical example is given by the set of \emph{bounded} linear operators $\mathcal B(\hilbert)$ on a Hilbert space $\hilbert$, where the norm is the usual operator norm given by the square root of
\[
\|A\|^2 = \sup_{\psi \in \hilbert} \frac{\bra \psi A^\dagger A\ket \psi}{\braket \psi \psi}.
\]
We see that the requirement of a finite norm eliminates unbounded operators such as most quantum mechanical observables (position, momentum, energy). However this is not a problem , since, as seen above, the interpretational content of these observables can be replaced by elementary \emph{effects} (in this case projectors) which are always bounded. We will come back to this point later when we need to talk about infinite-dimensional Hilbert spaces.

The above axioms can be justified by the fact that they precisely capture the nature of algebras of linear operators:
\begin{theo}
Any $C^*$-Algebra is a norm-closed $*$-subalgebra of $\mathcal{B}(\hilbert)$ for some Hilbert space $\hilbert$.
\end{theo}
The proof is beyond the scope of this document. Instead, we will focus on concrete $*$-algebras of matrices.


\subsection{Structure of matrix algebras}

In the following we would like to characterize the general structure of the ``matrix algebras'', namely the $*$-subalgebras of the algebra of complex matrices of a given dimension. Because of the finite-dimensionality of these spaces, these subalgebras are also automatically $C^*$-algebras with the operator norm $\|A\|$ defined by
\[
\|A\|^2 = \max_{\psi} \frac{\bra \psi A^\dagger A \ket \psi}{\braket \psi \psi}. 
\]
However, contrary to the infinite-dimensional case, we will not need to make use of the norm for characterising these algebras.


The smallest such algebra that one can build is the algebra ${\rm alg}(A)$ \emph{generated} by a single self-adjoint matrix $A$, i.e., all linear combinations of (non-zero) natural powers of $A$. This algebra is commutative and unital, and can be characterised as follows:
\begin{theo}\label{smallestalg}
If $A$ is a self-adjoint matrix with spectral decomposition $A = \sum_i a_i P_i$ (where $a_i \neq 0$ are distinct eigenvalues of $A$, and $P_1, \cdots, P_m$ are orthogonal projectors), then 
\[
{\rm alg}(A) = {\rm span}(P_1,\dots,P_m)
\]
namely, the $*$-algebra generated by $A$ is equal to the span of the spectral projectors of $A$. Explicitly,
\[
P_j = \prod_{i\neq j}\frac{A-a_i{\one_{A}}}{a_j-a_i},
\]
where $\one_{A} = A^0 =\sum_{i=1}^m P_i$ is an identity for the algebra.  
\end{theo}
\proof
We used the notation 
\[
{\rm alg}(A) = \Bigl\{{ \sum_{n=1}^N c_n A^n | c_n\in\mathbb{C}, N \in \mathbb N }\Bigr\}
\quad\text{ and }\quad
{\rm span}(P_1,\dots,P_m) = \Bigl\{{\sum_{i=1}^m c_i P_i| c_i\in\mathbb{C} }\Bigr\}.
\]
It is clear that any power of $A$ is inside the space spanned by the projectors $P_i$, since $A^n = \sum_i a_i^n P_i$. 
For the converse, to show that $\sum_i b_i P_i$ is inside ${\rm alg}(A)$, we have to find coefficients $c_1,c_2,\dots,c_N$ (for some $N$) such that $\sum_n c_n A^n = \sum_{ni} c_n a_i^n P_i = \sum_i b_i P_i$, i.e., $\sum_n c_n a_i^n = b_i$ for all $i$. This can alyways be solved because the matrix $\{a_i^n\}_{in}$, for $i,n=1,\dots, m$, is invertible. 
\qed

This result has a natural generalisation to arbitrary matrix algebras:
\begin{theo}
 \label{spanproj}
Any $*$-algebra $\mathscr A$ of complex matrices is unital, and is equal to the span of the projectors it contains:
\begin{equation}
\label{spanproj}
\mathcal A = {\rm span}\{ P \in \mathcal A : P^\dagger P = P\}.
\end{equation}
\end{theo}
\proof
If we consider an arbitrary element $B\in\mathscr{A}$ we can write it as combination of a hermitian and anti-hermitian part, i.e.
\[
B=\underbrace{\frac{B+B^*}{2}}_{=:\re(B)}+i\underbrace{\frac{B-B^*}{2i}}_{=:\im(B)}.
\]
$\re(B)$ and $\im(B)$ are both self-adjoint by definition, and manifestly inside of the same $*$-algebra $\mathscr A$. Moreover, theorem~\ref{smallestalg} showed that the spectral projectors of these matrices are also within $\mathscr A$, since it must contain ${\rm alg}(\re(B))$ and ${\rm alg}(\im(B))$. Hence if we write their spectral decompositions as $\re(B)=\sum_ib_i P_i$ and $\im(B)=\sum_ib_i Q_i$ then we obtain
\[
B = \sum_i a_i P_i + i\sum_j b_j Q_j,
\]
where $P_i,Q_j \in \mathscr A$. This proves Equ.~\eqref{spanproj}. 

By decomposing in such a way each element of a basis of $\mathscr A$, we can build a finite set of projectors $T_i$, $i=1,\dots,n$ which span $\mathscr A$. 
From theorem~\ref{smallestalg}, we know that the projector $P := (\sum_i T_i)^0$ on the range of $\sum_i T_i$ is also inside $\mathscr A$, and satisfies $P (\sum_i T_i) = \sum_i T_i$, which implies $\sum_i (\one - P) T_i (\one - P) = 0$. But this is a sum of positive operators, therefore each one of them must be zero: $(\one - P) T_i (\one - P) = 0$ for all $i$, as well as their square roots $T_i (\one - P) = (\one - P) T_i = 0$ (which can be proven by taking the expectation value of the previous expression in an aribtrary vector). It follows that $T_i P = P T_i = T_i$ for all $i$. Since the matrices $T_i$ span $\mathscr A$, this shows that $P \in \mathscr A$ is the identity of $\mathscr A$.  
\qed

\vspace{0.5cm}
Note that Equ.~\eqref{spanproj} is also true in any $*$-algebras of operators on a Hilbert space, even an infinite-dimensional one, provided that the algebra is closed in the \emph{weak operator topology}, i.e., that if a sequence of operators $A_n$ are such that $\bra \psi A_n \ket \psi$ converges to $\bra \psi A \ket \psi$ for all vectors $\ket \psi$ then $A$ is also inside the algebra. These algebras are special types of C$^*$-algebras called \emph{von-Neumann algebras}. The span of the projectors must then also be closed in that topology.
There are, however, many non-trivial C$^*$-algebras which contain no projector besides the identity and zero elements.

The unital algebra ${\rm alg}(A)$ generated by a single self-adjoint matrix $A$ is the prototype of a commuting matrix algebra. Indeed, all commuting algebras are of this form:
\begin{theo}\label{th:abelianalg}
Any commuting $*$-algebra $\mathcal A$ of complex matrices is of the form
\[
\mathcal A = {\rm alg}(A) = {\rm span}(P_1,\dots,P_n),
\]
where $A$ is a self-adjoint matrix and $P_1, \dots, P_n$, are mutually orthogonal projectors, and $n$ is the dimension of $\mathcal A$ (as a linear space).
\end{theo}
\proof
Consider a linear basis $B_1,\dots,B_n$ of $\mathcal A$. By decomposing each $B_j$ into a real and an imaginary part, one obtains a family $A_1,\dots,A_{m}$, where $m = 2n$, of self-adjoint operators spanning $\mathcal A$. Now consider the spectral decomposition of each of these operators: $A_j = \sum_{i=1}^{n_j} a^j_i Q^j_i$. Since the algebra is commutative, these projectors also commute with each other. This implies in particular that the product of any two such projectors is also a projector. Indeed, if $P$ and $Q$ are two commuting projectors, then $(PQ)^\dagger (PQ) = QPPQ = QPQ = PQQ = PQ$.

Also, if $P$ and $Q$ are two commuting projectors, then $R := P+Q-PQ$ is also a projectors which is such that $RP = P$ and $RQ = Q$. This means that it projects on the union of the ranges of $P$ and $Q$. In this way we can build a projector $\one_{\mathcal A} \in \mathcal A$ which projects on the union of the projectors $Q^j_i$ for all $i$ and $j$. This is an identity element for $\mathcal A$. We can use it to complete each spectral decomposition by a new element $Q_0^j = \one_{\mathcal A} - \sum_{i=1}^{n_j} Q_i^j$.

Let's consider the projectors $P_I = Q_{I_1}^1 \cdots Q_{I_m}^m$, where $I_i \in \{0,1,\dots,\}$. They are all orthogonal to each other. Indeed, if $I$ and $J$ differ by a single entry, then $P_I$ and $P_J$ contains in their product two orthogonal projectors, which, given that they all commute, implies $P_I P_J = 0$. Moreover, the span of these operators include also the projectors $Q_j^i$, and hence span the algebra. Hence we have obtained a complete family of orthogonal projectors $P_I$ spanning $\mathcal A$. Observe, however, that most of them are typically equal to zero. Indeed, due to their orthogonality, they are linearly independent, and hence only $n$ of them can be non-zero. Let $P_1, \dots, P_n$ be those nonzero projectors.  
If we then pick a distinct real number $a_i$ for each $i$, then the matrix $A = \sum_i a_i P_i$ generates the whole of $\mathcal A$ (since each $P_i$ is inside ${\rm alg}(A) \subseteq \mathcal A$). 
\qed

Now that we have understood the form of commutative matrix algebras, we gain a first handle on generical matrix algebras by considering a commutative algebra that they contain; their \emph{center}:
\begin{mydef}
Let $\mathscr{A}$ be a $*$-algebra. The \textbf{center} of $\mathscr{A}$ is the commutative algebra
\begin{equation}
\mathcal{Z}(\mathscr{A})\equiv\left\{A\in\mathscr{A} \ : \ \left[A,B\right]=0, \ \forall B \in \algebra  \right\}.
\end{equation}
\end{mydef}
Hence the center $\mathcal{Z}(\mathscr{A})$ is made of those elements of $\mathcal A$ which commute with all the other elements of $\mathcal A$. Clearly, $\mathcal{Z}(\algebra)$ is a $*$-algebra: If we take $A,B\in\mathcal{Z}(\algebra)$ we see that also $AB\in\mathcal{Z}(\algebra)$. Indeed, for all $C \in \algebra$,
$
ABC = ACB = CAB.
$
Moreover, $\mathcal{Z}(\mathscr{A})$ is clearly commutative (abelian), i.e., for all $A,B$ in $\mathcal{Z}(\mathscr{A})$, $[A,B]=0$.

The structure of the center $\mathcal{Z}(\mathscr{A})$ gives us already important information about the structure of a general matrix algebra $\mathscr{A}$. 
In order to express the resulting property, however, we need to introduce the concept of direct sum $\oplus$ for algebras. The algebraic direct sum $\oplus$ can be defined concretely as follows: given two square matrices $A$ and $B$, respectively of sizes $n \times n$ and $m \times m$, we define the matrix $A \oplus B$ of size $nm \times nm$ as
\[
A \oplus B = \begin{pmatrix} A & 0 \\ 0 & B \end{pmatrix}
\]
where the $0$'s represent rectangular matrices of the apropriate sizes with only zero components. It should be clear how this generalises for the direct sum of more than two matrices. 

This direct sum could also be defined algebraically also on general abstract algebras $\mathscr A$ and $\mathscr B$: $\mathscr A \oplus \mathscr B$ is the direct sum of $\mathscr A$ and $\mathscr B$ as linear spaces equipped with the product defined by 
\[
(A \oplus B) (A' \oplus B') := (AA') \oplus (BB').
\]
As block matrices, this is simply
\[
\begin{pmatrix} A & 0 \\ 0 & B \end{pmatrix} \begin{pmatrix} A' & 0 \\ 0 & B' \end{pmatrix} 
= \begin{pmatrix} AA' & 0 \\ 0 & BB' \end{pmatrix}.
\]
With the help of this concept, we obtain the first importance piece of information about general matrix algebras:
\begin{lemma}\label{lem:directsum}
For any $*$-algebra $\mathcal A$ of complex matrices, there is a unitary matrix $U$, and smaller matrix algerbas $\mathcal A_i$ with $\mathcal Z(\mathcal A_i) = \mathcal C \one$, such that
\[
U \mathcal A U^\dagger = \mathscr A_1 \oplus \dots \oplus \mathscr A_n \oplus 0.
\]
\end{lemma}
\proof
Theorem \ref{th:abelianalg} tells us that there exists a minimal set of orthogonal projectors $\{P_i\}_{i=1}^n$ such that $\mathcal{Z}(\mathscr{A})={\rm span}(P_1,...,P_n)$. Note that the sum of all projectors is the identity of $\algebra$, i.e.\ $\sum_iP_i=\one _{\algebra}$. This matrix $\one _{\algebra}$ should not be confused with the identity matrix: it ould be in general any arbitrary projector. 

Due to the fact that the $P_i$ commute with every element of the algebra, we obtain the following decomposition of any element $A\in\algebra$:
\begin{equation}
A=\one _{\algebra}A=\left(\sum_iP_i\right)A=\sum_i P_i P_i A = \sum_i P_i A P_i \simeq A_1 \oplus A_2 \oplus \dots \oplus A_n \oplus 0,
\end{equation}
for some smaller matrices $A_i$ whose dimensions sum to the total dimension of the projector $\one_{\mathscr{A}}$. The symbol $\simeq$ means that the two matrices are equal up to a change of orthonormal basis.


In this basis, we have,
\begin{align*}
\one_{\mathscr{A}} &\simeq \one_{d_1} \oplus \dots \oplus \one_{d_n} \oplus 0_{d_{n+1}}\\
\end{align*}
where $\one_{d_i}$ denotes the identity matrix of dimension $d_i$, where $d_i = \tr P_i$, and $0_{d_{n+1}}$ is the zero square matrix of dimension 
\[
d_{n+1} := d - \sum_{i=1}^n d_i
\]
where $d$ is the dimension of the original matrices in $\mathscr{A}$, such as $A$. This is just the identity matrix accept for a few trailing zeros on the diagonal ($d_{n+1}$ of them). 

Also, the smaller matrices $A_i$ are defined as representing the non-trivial parts of the matrices $P_i A P_i$ in the new basis:
\[
P_i A P_i \simeq 0_{d_1+\dots d_{i-1}} \oplus A_i \oplus  0_{d_{i+1} + \dots + d_{n+1}}.
\]

Hence, in this special basis we have, in block matrix notation,
\[
A \simeq
\begin{pmatrix} 
A_1 & & & & \\
& A_2 & & & \\
& & \ddots & & \\
& & & A_n & \\
& & & & 0\\
\end{pmatrix} \equiv A_1 \oplus A_2 \oplus \dots \oplus A_n \oplus 0.
\]
where the blank spots are filled with zeroes, and $0$ represents the zero square matrix of dimension $d_{n+1}$. 

Because the projectors $P_i$ only depend on the algebra $\mathscr{A}$, not on the chosen element $A$, all elements of $\mathscr{A}$ have this particular form when expressed on that specific basis (but for different values of the small matrices $A_i$). In fact, it is easy to see that the allowed values of the matrices $A_i$ must also form a $*$-algebra that we call $\mathscr{A}_i$, so that we obtain:
\[
\mathscr{A} \simeq \mathscr A_1 \oplus \dots \oplus \mathscr A_n \oplus 0.
\]
\qed

In order to completely elucidate the structure of $\mathscr{A}$, we still need to understand the structures of the smaller algebras $\mathscr A_i$. What distinguishes them from a general matrix algebra like $\mathscr A$ is the fact that their center is ``trivial'', i.e., it consists of all multiples of the identity: $\mathcal Z(\mathscr A_i) = \mathbb C \one$.
Indeed, if there were other elements in the center of $\algebra_i$, there would be a finer decomposition of the minimal set of projectors of the center. Such a matrix algebra with a trivial center is called a \emph{factor}. 

We now show that such a matrix algebra factor always consists of all matrices of the form $\one \otimes A$ in some basis:
\begin{lemma}\label{lem:simple}
Let $\mathscr{A}$ be a $*$-algebra of $d$-dimensional complex matrices, with trivial center (i.e., a factor): 
\(
\mathcal Z(\mathscr A) = \mathbb C \one.
\)
Then, there is a unitary matrix $U$ such that
\begin{equation}
U \mathcal A U^\dagger = \one_m \otimes \mathcal{B}(\mathbb{C}^n).
\end{equation}
with $mn = d$.
\end{lemma}
\proof
Our proof is adapted from Ref.~\cite{takesaki}.
Given any $A \in \mathcal A$ such that $A = A^\dagger \neq 0$, consider the space $\mathcal I_A := {\rm span} \{ X A Y : X,Y \in \mathcal A\}$. 
It is clear that $\mathcal I_A$ is closed under multiplication and under the $\dagger$ operation, and hence forms a matrix algebra. From theorem~\ref{spanproj} we know that $\mathcal I_A$ contains an identity $P \in \mathcal I_A$. This implies that for all $B \in \mathcal A$, $PB = (PB)P = P(BP) = BP$ since $BP \in \mathcal I_A$. Therefore, $P \in \mathcal Z(\mathcal A)$. But since $\mathcal A$ is a factor this implies $P = \one \in \mathcal I_A$. 
In other words, $\one = \sum_i X_i A Y_i$ for some $X_i, Y_i \in \mathcal A$.

It follows that for all $A,B \in \mathcal A$, $A = A^\dagger \neq 0$, $B = \sum_i B X_i A Y_i$. If $B \neq 0$, this implies that at least one of the terms in the sum must be nonzero: $B X_i A Y_i \neq 0$, which implies $B X_i A \neq 0$.

We have therefore shown that for all $A,B \neq 0 \in \mathcal A$ with $A = A^\dagger$, there exists $X \in \mathcal A$ such that $B X A \neq 0$. 
This is the property we need in what follows. 

Now let us consider a \emph{maximal} commutative $*$-subalgebra $\mathcal C$ of $\mathcal A$. The maximality means that if $B \in \mathcal A$ is such that $[B,A] = 0$ for all $A \in \mathcal C$, then $B \in \mathcal C$. We can build one as follows: we start with $\mathcal C_0 = \mathbb C \one$, then recursively build $\mathcal C_{i+1}$ as the span of $\mathcal C_i$ and any linearly independent self-adjoint element which commutes with everything in $\mathcal C_i$. The final algebra $\mathcal C$ is clearly commutative, and has the desired property, because if $B$ commutes with all elements of $\mathcal C$, then $B^\dagger$ also does, as well as $\re B = \frac 1 2 (B + B^\dagger)$ and $\im B = \frac 1 {2i}(B - B^\dagger)$ which are self-adjoint and therefore in $\mathcal C$. It follows that $B = \re B + i\, \im B \in \mathcal C$.

This exists because we can build it by progressively adding any normal element $B$ commuting with all elements of a given commutative subalgebra.

From theorem \ref{th:abelianalg}, we know that $\mathcal C = {\rm span}(P_1,\dots,P_n)$, where $P_i$ are a complete family of orthogonal projectors. 
We have just shown above that for every pair $i,j$, there exists a $X_{ij} \in \mathcal A$ such that 
\[
F_{ij} := P_i X_{ij} P_j \neq 0.  
\]
Observe that $F_{ij}^\dagger F_{ij}$ commutes with every elements of $\mathcal C$ and is therefore contained in $\mathcal C$ due to its maximality. This means that it is a linear combination of the projectors $P_k$, but its product with $P_k$, $k \neq j$ is zero, hence 
\[
F_{ij}^\dagger F_{ij} = P_j X_{ij}^\dagger P_i X_{ij} P_j = \alpha_{ij} P_j
\]
for some $\alpha_{ij} > 0$. In particular, this implies that all the $P_i$'s project on spaces of same dimensions $m$ since $F_{ij}^\dagger F_{ij}$ above maps all vectors in $P_j \mathbb C^d$ via $P_i \mathbb C^d$ back to $P_j \mathbb C^d$ without sending any to zero. Therefore, we can decompose $\mathbb C^d$ into $\mathbb C^m \otimes C^n$ with an orthonormal basis $\ket i \otimes \ket j$ such that
\[
P_i = \sum_{j=1}^m \proj{j} \otimes \proj i = \one \otimes \proj i.
\]
This yields $F_{ij} = Y_{ij} \otimes \ketbra ij$, where $Y_{ij} = (\one \otimes \bra i) X_{ij} (\one \otimes \ket j)$. 

But then $F_{ij}^\dagger F_{ij} = Y^\dagger_{ij} Y_{ij} \otimes \proj i = \alpha_{ij} \one \otimes \proj i$, which implies $Y^\dagger_{ij} Y_{ij} = \alpha_{ij} \one$, and hence
\[
F_{ij} = \sqrt{\alpha}_{ij} U_{ij} \otimes \ketbra ij
\]
for some unitary matrices $U_{ij}$.
For convenience, let us also define $F_{ii} = P_i = \one \otimes \proj i$.

For any $A \in \mathcal A$, consider the operator $P_i A P_j X_{ij}^\dagger P_i$. This operator commutes with all projectors $P_i$, and hence belongs to $\mathcal C$, which implies that $P_i A P_j X_{ij}^\dagger P_i \propto P_i$. It follows that 
\[
P_i A P_j = P_i A P_j P_j \propto P_i A P_j F_{ji}^\dagger F_{ji} \propto P_i A P_j X_{ij}^\dagger P_i X_{ij} P_j \propto P_i X_{ij} P_j = F_{ij}.
\]
But also $A = \sum_{ij} P_i A P_j$, therefore $A$ is a linear combination of the operators $F_{ij}$.

For $A = F_{ij} F_{jk}$, we obtain that $F_{ij} F_{jk} \propto F_{ik}$, which implies $U_{ij} U_{jk} \propto U_{ik}$. Similarly, $U_{ij_1} U_{j_1 j_2} \dots U_{j_n} U_{j_n i} \propto \one$. In particular, $U_{ji} \propto U_{ij}^\dagger$.

We still need a change of basis to put the generators $F_{ij}$ of $\mathscr A$ in the form that we want. It is provided by
\[
U = \sum_i U_{ij} \otimes \proj i
\]
where the choice of $j$ is arbitrary.
Indeed, we obtain
\[
U^\dagger F_{ik} U \; \propto \; U_{ij}^\dagger U_{ik} U_{kj}  \otimes \ketbra i k 
\; \propto \;\one \otimes \ketbra i k.
\]
\qed

It is now straightforward to combine Lemma \ref{lem:directsum} and \ref{lem:simple} in order to completely elucidate the structure of general matrix algebras:
\begin{theo}\label{th:algebrastructure}
For any $*$-algebra subalgebra $\mathcal A$ of $\mathcal{B}(\mathbb{C}^d)$, there is a unitary matrix $U$ such that
\begin{equation}
U\algebra U^\dagger = \left[{ \bigoplus_{i=1}^N \one _{m_i}\otimes M_{n_i}(\mathbb{C})}\right] \oplus 0_{d_0}
\end{equation}
where $\sum_{i=1}^{N}n_im_i + d_0 =  d$.
\end{theo}
Equivalently, this means that the elements of $A$ are precisely those matrices of the form
\begin{equation}
A = U^\dagger
\begin{pmatrix} 
\one_{m_1} \otimes A_1 & & & \\
& \ddots & & \\
& & \one_{m_N} \otimes A_N & \\
& & & 0\\
\end{pmatrix}
U
\end{equation}
for any set of square matrices $A_i$ of size $n_i$. 

In this notation, the center $\mathcal{Z}(\mathscr{A})$ of $\mathcal A$ is made of all matrices of the form
\begin{equation}
A = U^\dagger 
\begin{pmatrix} 
\alpha_1 \one_{m_1} \otimes \one_{n_1} & & & \\
& \ddots & & \\
& & \alpha_N \one_{m_N} \otimes \one_{n_N} & \\
& & & 0\\
\end{pmatrix}
U
\end{equation}
for any family of complex numbers $\alpha_1,\dots\alpha_N$.


\subsection{Hybrid quantum/classical systems}


Theorem~\ref{th:algebrastructure} allows us to understand the type of physical system represented by a generic matrix algebra $\mathcal A$, in terms of the state/effect formalism.
For short, we write 
\begin{equation}
\label{equ:hybrideffect}
A = \bigoplus_i \one \otimes A_i
\end{equation}
for generic element of $\mathcal A$, according to the decomposition given by Theorem~\ref{th:algebrastructure}. For conciseness, in this notation we leave implicit the possible trailing zero block. Also, one must remember that this is block-diagonal only in a basis which may not be the canonical one.

 It is easy to see that $A^\dagger = A$ if and only if $A_i^\dagger = A_i$ for each $i$. Moreover, by writing $A$ in diagonal form, we immediatly see that the eigenvalues of $A$ are all in the interval $[0,1]$ if and only if that is the case also for each $A_i$. Therefore, we conclude that the effects $0 \le A \le \one$ of $\mathcal A$ are precisely the operators of the form $\bigoplus_i \one \otimes A_i$ where $0 \le A_i \le \one$ for each $i$.


From theorem \ref{th:riesz}, we know that the states can be represented by matrices $R$ as the functionals
\[
A \mapsto \tr(RA)
\]
on effects $A$. But since the set of effects is restricted, two matrices $R$ and $R'$ may actually represent the same functional, i.e., $\tr(R A) = \tr(R' A)$ for all $A \in \mathcal A$. 

Let us find a special matrix $R$ giving a unique representation of the functional.
Let $P_1,\dots,P_N$ be the projectors spanning the center of $\mathcal A$. We know that the elements of $\mathcal A$ satisfy $A = \sum_i P_i A P_i$. Therefore, 
\[
\tr(R A) = \sum_i \tr(R P_i A P_i) = \sum_i \tr(P_i R P_i A).  
\]
This means that the functional is uniquely specified by the matrix
\[
R' = \sum_i P_i R P_i = \bigoplus_i R_i.
\]
There is however still some ambiguity left on the matrices $R_i$ since
\[
\sum_i \tr\bigl({ R_i (\one \otimes A_i) }\bigr) = \sum_i \tr\bigl({ (\one \otimes \tr_i'(R_i)) (\one \otimes A_i) }\bigr).
\]
We obtain that the same functional is uniquely represented by a matrix of the form
\[
R = \bigoplus_i \one \otimes R_i,
\]
namely an element of the algebra $\mathcal A$ itself. 

Using the same argument, as for the effects, this matrix $R$ is positive if and only if $R_i \ge 0$ for all $i$. The last condition for $R$ to represent a \emph{state} is the normalisation condition $\tr R = 1$. Since $R_i$ is positive, we can write $R_i = r_i \rho_i$ where $\rho_i$ is a normalised density matrix and $r_i := \tr R_i$. The normalisation condition then says
\[
\tr R = \sum_i m_i r_i = 1.
\]
Let us write $p_i = m_i r_i$. These numbers form a probability distribution, and we have
\begin{equation}
\label{hybrid}
R = \bigoplus_i \,p_i\, \frac{\one_i}{m_i} \otimes \rho_i,
\end{equation}
which is a unique representation for a state of the theory defined by the algebra $\mathcal A$. 
 
We make two observations. The first is that $R$ is also a density matrix on the quantum system defined by the full matrix algebra of which $\mathcal A$ is a subalgebra. Therefore, restricting the quantum effects to those of the subalgebra $\mathcal A$ is equivalent to restricting the density matrices also to $\mathcal A$.

The second observation, is that a state of the system defined by $\mathcal A$ is represented by a probability distribution $p_1,\dots,p_N$, hence a \textit{classical system}, together with a set of quantum states $\rho_1,\dots,\rho_N$. 
Hence it is a sort of \textit{hybrid quantum/classical system}. 
This interpretation is justified by the fact that these states characterise knowledge about the outcome of a
procedure where, say, a different quantum state is prepared depending on the value of a classical variable. 

Moreover, it is clear that if $N=1$, then $\mathcal A$ is just a quantum system, while if the dimension of each factor $\mathcal A_i$ is equal to $1$, then this is just a classical system.

A particular example of such a hybrid system is one which consists in a classical system next to a quantum one. 
Within the algebraic framework, we can compose systems of different type. For this purpose, we generalise the tensor product introduced in Section~\ref{reduced:state} in the context of quantum systems to systems defined by arbitrary algebras, $\algebra_1,\algebra_2 \rightarrow \algebra_1\otimes\algebra_2$. Algebras are vector spaces, for which the tensor product is defined in the usual way. In addition, we must defined how the product and $*$ operations, which are specific to the algebraic structure, are built on the composed systems.
Specifically, we demand the following properties for the multiplicative structure:
\begin{align} 
(A_1\otimes A_2)(B_1\otimes B_2) &= A_1B_1\otimes A_2B_2\\
(A_1\otimes A_2)^* &= A_1^*\otimes A_2^*
\end{align}
Importantly, there are canonical embeddings of a single algebra into the tensored space, which allow one to think of each algerba $\algebra_i$ as a subalgebra of $\algebra_1\otimes \algebra_2$, via the homomorphisms
\begin{align*}
\algebra_1 &\rightarrow \algebra_1\otimes\algebra_2
& \algebra_2 &\rightarrow \algebra_1\otimes\algebra_2\\
A &\mapsto A\otimes \one   
&  A &\mapsto \one \otimes A
\end{align*}
 
In case of matrix algebras, which are $*$-subalgebras of full matrix algebras, the tensor product is precisely the same as the quantum one, which is also called the Kronecker product, e.g.,
\begin{equation*}
\begin{pmatrix}
             a_{11} & \dots & a_{1n}\\
             \vdots & \ddots & \vdots \\
              a_{n1} & \dots & a_{nn}\\
\end{pmatrix}
\otimes B =
\begin{pmatrix}
             a_{11} B & \dots & a_{1n} B\\
             \vdots & \ddots & \vdots \\
              a_{n1} B & \dots & a_{nn} B\\
\end{pmatrix},
\end{equation*}
where $a_{ij}$ are the components of a matrix, and $B$ represents another matrix. 

For instance, if we compose two finite dimensional quantum systems $\algebra_1=\mathcal{B}(\hilbert_1)$ and $\algebra_2=\mathcal{B}(\hilbert_2)$, the tensor product is isomorphic to bounded operators on the tensor products of the hilbert spaces $\algebra_1\otimes\algebra_2 \equiv \mathcal{B}(\hilbert_1\otimes \hilbert_2)$. 

Recall that a classical system associated with the finite set of pure states (sample space) $\Omega$ corresponds to the commutative algebra of functions $\mathcal A = \funs(\Omega)$ on $\Omega$, namely the space of vectors with elements indexed by elements of $\Omega$, equipped with the component-wise product.
Taking the tensor product of two classical systems $\algebra_1=\funs(\Omega_1)$ and $\algebra_2=\funs(\Omega_2)$ results in the set of functions on the \emph{cartesian products} of the sample spaces: $\algebra_1\otimes\algebra_2 \equiv \funs(\Omega_1\times\Omega_2)$. Alternatively, one may also view those commutative algebras as sets of diagonal matrices, with diagonal elements index by $\Omega$. The Kronecker product then gives the same result. 

A natural \emph{hybrid} system then is obtained by considering the composition of a classical system $\algebra_1=\funs(\Omega)$ and quantum system $\algebra_2=\mathcal{B}(\hilbert)$. The easiest way to see what happens is, again, to represent $\funs(\Omega)$ as a set of diagonal matrices. Using the Kronecker product, one then find that the tensor product yields a block-diagonal algebra
\[
\funs(\Omega)\otimes \mathcal{B}(\hilbert)\;\; \simeq \;\; \mathcal{B}(\hilbert) \oplus \dots \oplus\mathcal{B}(\hilbert),
\]
where each factor $\mathcal B(\hilbert)$ on the right hand side of the equation is indexed by an element of $\Omega$.


For instance, a completely uncorrelated state on $\algebra_1\otimes \algebra_2$ has the form 
\begin{align*}
\{p_i\}_{i=1}^{N}\otimes\rho \equiv
\begin{pmatrix}
            p_1 & & \\
             & \ddots &  \\
             & & p_N\\
\end{pmatrix}
\otimes \rho=
\begin{pmatrix}
            p_1 \rho & & \\
             & \ddots &  \\
             & & p_N \rho\\
\end{pmatrix}
\equiv \sum_i p_i \proj i \otimes \rho
\end{align*}
where the orthogonal vector states $\ket i$ just serves for the matrix representation of the classical state $\{p_i\}_{i=1}^N \equiv \sum_i p_i \proj i$.

Similarly, a general correlated classical/quantum state has the form
\begin{align*}
\begin{pmatrix}
            p_1 \rho_1 & & \\
             & \ddots &  \\
             & & p_N \rho_N\\
\end{pmatrix}
&=\sum_{i=1}^N p_i\proj{i}\otimes \rho_i.
\end{align*}
This is a special case of the general expression Equ.~\eqref{hybrid}, where $\proj i$ is simply the identity operator on a \emph{one-dimensional} factor, and where each state $\rho_i$ leaves in a Hilbert space of same dimension.

%% file: channels.tex
\section{Channels}

We want to identify the most general way of representing a transfer of information from a system represented by some algebra $\algebra_1$ to another system represented by $\algebra_2$. This can be done in two ways: either one maps states to states (Schr\"odinger picture), or effects to effects (Heisenberg picture). In infinite dimensions, the Heisenberg picture is more general. However, we begin with the Schr\"odinger picture because it is somewhat more intuitive. 

Mathematically, a map preserving \emph{all} the structure of $\algebra_1$ is given by the following definition:
\begin{mydef}[$*$-homomorphism]
Let $\algebra_1$ and $\algebra_2$ be $*$-algebras. A map $\phi:\algebra_1 \rightarrow \algebra_2$ is referred to as \textbf{$*$-homomorphism} (or $*$-algebra morphism) if it satisfies
\begin{enumerate}
	\item $\phi$ is linear
	\item $\phi(AB)=\phi(A)\phi(B)$
	\item $\phi(A^*)=\phi(A)^*$
\end{enumerate}
\end{mydef}
However, these conditions are too strong for our purpose. Indeed, we only want to preserve the state/effect structure. This structure does rely on the algebra product, but only through the partial order defined via the concept of positivity. Indeed, recall that positivity is defined using the product (Definition~\ref{def:positive}). 
But, as we will see, the requirement that a map preserves this partial order is much weaker than requiring that it preserves the product structure.

\subsection{Schrödinger Picture}



A map $\channel$ on density matrices should preserves convex combinations of states. Indeed, 
since the state $\sum_i p_i \rho_i$---where $\{p_i\}_{i=1}^n$ is a probability distribution and $\rho_i$ are density matrices---represents the belief that the state is $\rho_i$ with probability $p_i$, its image under the map $\channel$ should be
\[
\channel \bigl( \sum_i p_i \rho_i \bigr) = \sum_i p_i \channel(\rho_i),
\]
namely, the image of $\rho$ under $\channel$ ought to be state $\channel(\rho_i)$ with probability $p_i$. Since the set of density matrices for the system represented by $\mathcal A_1$ span the whole of $\mathcal A_1$, this map can be naturally extended to a linear map 
\[
\channel: \algebra_1 \rightarrow \algebra_2
\]

Moreover, for the image of any density matrix to also be a density matrix, the map $\channel$ must send positive matrices to positive matrices (according to which we simply say that $\channel$ is itself \emph{positive}), and it must also preserve the trace of matrices:
\[
X \ge 0 \;\;\Rightarrow\;\; \channel(X) \ge 0,
\]
and
\[
\tr(\channel(X)) = \tr(X).
\]

But there is an extra more subtle consequence of the positivity condition. Namely, $\channel$ should also respect all those conditions if we see as part of a larger system. That is, if we add any system $\mathcal{B}(\hilbert)$ on which $\channel$ acts trivially. On the large system, this is represented by the map $(\channel \otimes id):\algebra_1\otimes \mathcal{B}(\hilbert)\rightarrow\algebra_2\otimes\mathcal{B}(\hilbert)$ defined by 
\[
(\channel \otimes id)(X \otimes Y) = \channel(X) \otimes Y.
\]
Classically, one would expect that the positivity of $\channel$ implies that of $\channel \otimes id$. Indeed, if the algebra $\mathcal A_1$ is commutative, then a general operator in $\mathcal A_1 \otimes \mathcal B(\hilbert)$ is of the form
\[
X = \sum_i \proj i \otimes X_i,
\]
where $X_i \in \mathcal B(\hilbert)$. Since the eigenvalues of $X$ are just those of the $X_i$'s, it is easy to see that $X \ge 0$ if and only if $X_i \ge 0$ for all $i$. We then have
\[
(\channel \otimes id)(X) = \sum_i \channel(\proj i) \otimes X_i.
\]
Since $\channel(\proj i)$ and $X_i$ are positive, so is the right hand side of this equation. This shows that, when $\mathcal A_1$ is commutative, and hence represents a classical system, the positivity of $\channel$ implies that of $\channel \otimes id$. 
However, the following example shows that this fails when $\mathcal A_1$ is quantum:
\begin{ex}[{Partial Transpose}]
Consider the transpose map $\channel(\rho)=\rho^\top$, where $\rho$ is state on $\hilbert_A$. 
Since the transposition leaves the eigenvalues of $\rho$ invariant, this is clearly a positive map (and trace preserving). However, consider the setting if we compose the Hilbert space with an ancillary space of the same dimension $d$: $\hilbert=\hilbert_A\otimes \hilbert_B$. We extend $\channel$ by the identity on this ancillary space and consider the action on the maximally entangled state $\ket{\Omega}=\frac{1}{\sqrt{n}}\sum_i\ket{i}_A\otimes\ket{i}_B$:
\begin{equation}
(\channel\otimes\one )(\proj{\Omega})=(\channel\otimes\one )\left(\frac{1}{d}\sum_{ij}\ketbra ij \otimes \ketbra ij \right)=\frac{1}{d}\sum_{ij}\ketbra ji \otimes \ketbra ij =:\tilde{\rho}
\end{equation}
But the operator $\tilde{\rho}$ is not positive. Indeed, consider the case $d=2$, and the state $\ket \psi = \frac{1}{\sqrt{2}}(\ket{01}-\ket{10})$. A direct calculation shows that
\[
\tilde \rho \ket \psi = - \ket \psi,
\]
from which it follows that $\bra \psi \tilde \rho \ket \psi = -1$.
\end{ex}


We conclude that positivity is not a sufficient criterion for a map to fit into the state/effect formalism of quantum systems. We therefore need to extend this notion to that of \emph{complete positivity}:
\begin{mydef}[Complete Positivity]
A linear map $\channel:\algebra_1 \rightarrow \algebra_2$ is \emph{completely positive}, iff $\channel\otimes id_n: \algebra_1\otimes \mathcal{B}(\mathbb{C}^n) \rightarrow \algebra_2\otimes \mathcal{B}(\mathbb{C}^n)$ is positive for all n, i.e.
\begin{equation}
\forall n\in\mathbb{N}: \quad (\channel\otimes id_n)(X)\geq 0 \ \ \forall X\geq 0.
\end{equation}
\end{mydef}

This gives us all the tool to define a general channel in the Schr\"odinger picture:
\begin{mydef}[Channel] 
A channel represents a general transfer of information from a system represented by that algebra $\mathcal A_1$ to that represented by $\mathcal A_2$. When those are matrix algebras, it can be represented by an arbitrary linear, completely positive, trace-preserving (CPTP) map $\channel:\algebra_1\rightarrow \algebra_2$, meant to be applied to states (Schr\"odinger picture).
\end{mydef}


In what follows we give some simple examples of channels.

\begin{ex}[Homomorphisms]
A $*$-homomorphism $\phi: \mathcal A_1 \rightarrow \mathcal A_2$ is a channel if it is trace-preserving. In order to show this, we just need to show that it is completely positive. Moreover, since $\phi \otimes id$ is also a $*$-homomorphism, we only need to show that an arbitrary $*$-homomorphism like $\phi$ is positive. This can be done by using Definition~\ref{def:positive}, which says that $A \ge 0$ if and only if $A = \sum_i E_i^* E_i$. We then have
\[
\phi(A) = \sum_i \phi(E_i^*  E_i) = \sum_i \phi(E_i)^*  \phi(E_i).
\]
The right-hand side is manifestly positive, which completes the proof that $\phi$ is a channel.

An example of $*$-homomorphism is that induced by a \emph{unitary} operator $U$ (defined by the property $U^\dagger U = U U^\dagger  = \one$), through
\[
\phi(\rho) = U \rho U^\dagger .
\]
In fact, we will see below that when $\mathcal A_1 = \mathcal A_2$ is a matrix algebras, then a $*$-homomorphism is necessarily of this form.  

%
\end{ex}

\begin{ex}[Mixture of unitaries]

The set of channels between two given algebras is \emph{convex}. Indeed, if $\mathcal E_1$ and $\mathcal E_2$ are two channels, and $p$ a probability, then we can define the convex combination $\mathcal E = p \mathcal E_1 + (1-p) \mathcal E_2$ via
\[
\mathcal E(\rho) = p \mathcal E_1(\rho) + (1-p) \mathcal E_2(\rho).
\]
This map $\mathcal E$ is clearly trace-preserving. It is also easy to check that it is completely positive, because $\channel \otimes id$ is a convex combination of $\channel_1 \otimes id$ and $\channel_2 \otimes id$, and the convex combination of positive operators is still positive.

This gives us a first example of channels which are not homomorphisms, and hence go beyond unitary maps:
Let $\algebra_1=\algebra_2=\mathcal{B}(\hilbert)$ be the input and output algebra. Take $U_i:\hilbert \rightarrow \hilbert$ be a set of unitary linear operators and $\{p_i\}_{i=1}^N$. Define the map $\channel(\rho)=\sum_i p_i U_i\rho U_i^\dagger $. 
It is trace-preserving and completely positive.
This channels describes the situation in which the unitary $U_i$ is applied with probability $i$. Indeed, the resulting state is the convex combination of the possible outcomes, which is compatible with the interpretation of the convex combination of states. 
\end{ex}

\begin{ex}[Classical channels]
This definition of channel applies to any system defined by an algebra, not only quantum systems.
Let the input and output algebra of our system be classical $\algebra_1=\funs(\Omega_1), \algebra_2=\funs(\Omega_2)$. We have seen above that whenever $\algebra_1$ is classical (commutative) then positivity is sufficient. 
The linearity of $\channel$ makes it possible to write the action of $\channel$ on a probability distribution $\{p_i\}_{i\el\Omega_1}$ as
\[
\channel(\{p_i\})=\{q_i=\sum_j\pi_{ij}p_j\}_{i\el\Omega_2}
\]
The positivity of $\channel$ implies $\pi_{ij}\geq 0 , \ \forall i,j$. The trace-preserving property amounts to
\[
\sum_i q_i=\sum_{ij}\pi_{ij}p_j=\sum_j p_j \quad \forall \{p_j\}_{j=1}^N
\]
Choosing the particular distribution $p_j=\delta_{ij}$ we obtain $\sum_i \pi_{ij}=1 \ \forall j$. This means that a classical channel is represented by a \emph{stochastic} matrix $\pi$ which is the well-known concept of channel from usual probability theory.
\end{ex}

\subsection{Heisenberg Picture}

%
%

Consider the algebra of bounded operators on some finite dimensional Hilbert space $\mathcal{B}(\mathbb{C}^d)$, also known as $d$-by-$d$ complex matrices.  This is also a vector space, which we can equip with the Hilbert-Schmidt inner product 
\[
\langle A,B\rangle:=\Tr(A^\dagger  B).
\]
We have used this scalar product before, so as to combine a density matrix and an effect to obtain a probability. Indeed, the density matrix $\rho$ defines a state which associates the probability 
\[
p = \tr(\rho E) = \ave{\rho, E}
\]
to the truth value of the proposition defined by the effect $E$.

Now consider two Hilbert space that of Alice: $\hilbert_A$ and that of Bob: $\hilbert_B$.
A channel from Alice to Bob: $\channel : \mathcal{B}(\hilbert_A)\rightarrow\mathcal{B}(\hilbert_B)$ is in particular a linear map and hence has an adjoint $\channel^\dagger $ with respect to the Hilbert-Schmidt inner products on both algebras:
\[
\langle \channel(\rho),E\rangle = \langle \rho, \channel^\dagger  (E)\rangle
\]
or, explicitely,
\[
\tr(\channel(\rho) E) = \tr(\rho \,\channel^\dagger (E)).
\]
The quantity on the left hand side of those equations has a physical intepretation: it is the probability that the state $\channel(\rho)$ of Bob associates to the effect $E$ of Bob. But we see here that is also has a different interpretation from the point of view of Alice: it is the probability that Alice's state $\rho$ assigns to her effect $\channel^\dagger  (E)$.

It is important to note that the adjoint map $\channel^\dagger $ exchanges image and preimage, i.e.\ $\channel^\dagger   : \mathcal{B}(\hilbert_B)\rightarrow\mathcal{B}(\hilbert_A)$.


In this document, we refer to $\channel$ as the channel in the Schrödinger picture, which acts on states, and to $\channel^\dagger  $ as channel in the \emph{Heisenberg} picture, as it acts on effects. 

A familiar example is given by unitary channels. If $\channel(\rho) = U \rho U^\dagger $, then $\channel^\dagger (E) = U^\dagger  E U$. Indeed,
\[
\Tr(E\channel(\rho))=\Tr(E U\rho U^\dagger )=\Tr(U^\dagger EU\rho)=\Tr(\channel^\dagger  (E)\rho).
\]

We have already established the properties that the map $\channel$ needs to be a channel, but what do those properties correspond to in terms of $\channel^\dagger $? One could simply derive them directly from those of $\channel$. But, equivalently, one may also follow the same principle used to derived the properties of $\channel$, and demand that $\channel^\dagger $ preserves convex combinations of effects, which have the same ignorance interpretation as for convex combinations of states, and that it maps effects to effects, even when acting on part of a larger system. The first condition implies that $\channel^\dagger $ be linear. Since effects are defined in part by the fact that they are positive, the second condition implie that $\channel^\dagger $ be completely positive.

Hence, both $\channel$ and $\channel^\dagger $ are linear completely positive maps. The only difference is that whereas $\channel$ must be trace-preserving, $\channel^\dagger $ must be \emph{unital}, i.e., $\channel^\dagger (\one) = \one$. 


Indeed, for all operators $X$, we have
\[
 \Tr(\channel^\dagger  (\one )X)  = \Tr(\one \channel(X)) = \Tr(\channel(X)) =\tr(X) = \Tr(\one X) .
\]
Since this holds for all $X$, this implies
\[
\channel^\dagger (\one) = \one.
\]

This leads us to a more general definition of channel which holds in more general terms (including in the abstract framework of C$^*$-algebras):
\begin{mydef}[Channel, more general]
A channel transferring information from the system defined by the algebra $\mathcal A_1$ to that defined by the algebra $\mathcal A_2$ is represented by a linear, completely positive map
\[
\channel^\dagger : \mathcal A_2 \rightarrow \mathcal A_1.
\]
which is also unital: $\channel^\dagger (\one) = \one$, and meant to be applied to effects.
\end{mydef}
In order to avoid confusion, we will typically use the symbol $\dagger$ when using the Heisenberg picture, even if there is not corresponding Schr\"odinger representation. 

Again, it is important to note that in the Heisenberg representation, channels ``run backward'' in time. A consequence of this is that they must be composed in the opposite order from the Schr\"odinger representation, since
\[
(\channel_1\circ\channel_2)^\dagger  =\channel_1^\dagger  \circ\channel_2^\dagger 
\]

The reason this representation is preferred is that for a general C$^*$-Algebras $\algebra$, effects are elements of $\algebra$, but states are just linear functionals on $\mathcal A$. 
In fact, states can be seen themselves as channels from the system defined by the one-dimensional algebra $\mathbb C$ to that defined by $\mathcal A$, which, in the Heisenberg picture, is given by a linear completely positive unital map $\varphi: \mathcal A \rightarrow \mathbb C$. (This hints at the fact that we are secretly working in the category whose objects are C$^*$-algebras, and whose morphisms are unital CP maps).
In this language, the effect of $\channel^\dagger : \mathcal A_2 \rightarrow \mathcal A_1$ on the state $\varphi: \mathcal A_1 \rightarrow \mathbb C$ of the system $\mathcal A_1$ is given by composition: $\varphi$ is mapped to the new state $\varphi': \mathcal A_2 \rightarrow \mathbb C$ defined by
\[
\varphi' = \varphi \circ \channel^\dagger .
\]


\subsection{Observables as channels with classical output}

A type of channel which will play an important role in this lecture are those representing a transfer of information from any system to a classical system. 

For instance, such channels naturally represent generalised \emph{observables} of a system. Indeed, in the most general terms, an observation is a process by which information about the system being observed (represented by the algebra $\mathcal A$) is transferred to a measurement apparatus which, crucially, is a classical system. Let $\Omega$ be a set representing the relevant degrees of freedom of the classical measurement apparatus (say the possible positions of its ``dial''). 

\begin{figure}
	\centering
		\includegraphics[width=0.5\textwidth]{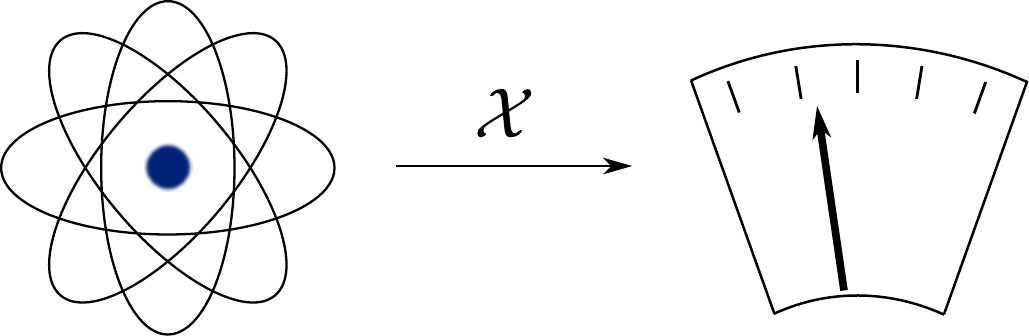}
		\vspace{0.5cm}
	\caption[Observables are quantum to classical channels]{An observable is a transfer of information from the system being observed to a pointer recording the result of a measurement.}
	\label{fig:QC-channel}
\end{figure}

Hence the following definition:
\begin{mydef}[Observables]
An \emph{observable} of a theory defined by the algebra $\mathcal A$ is any channel from that system to a classical system with pure states $\Omega$, represented in the Heisenberg picture by a unital CP map
\[
\qcmap^\dagger : \funs(\Omega) \rightarrow \mathcal A.
\]
The set $\Omega$ represents the possible \emph{values} of that observable. 
\end{mydef}
If this map has a Schr\"odinger representation 
\[
\qcmap: \mathcal A \rightarrow \funs(\Omega),
\]
then it maps a state $\rho$ of $\mathcal A$ to a probability distribution $\qcmap(\rho)$ over the possible set of values $\Omega$.

\begin{ex}[Observable associated with an effect]
Every quantum effect $E$ is naturally associated with an observable: the channel represented in the Schr\"odinger picture by the TPCP map $\qcmap_E: \mathcal B(\hilbert) \rightarrow \funs(\Omega)$, with $\Omega = \{{\textrm false},{\textrm true}\}$ and
\[
\qcmap_E(\rho) = \Tr(E\rho)\, \proj{\textrm true}+\Tr((\one -E)\rho)\, \proj{\textrm false},
\]
where we used the orthogonal vectors $\ket{\textrm true}$ and $\ket{\textrm false}$ to represent the two pure states of the classical apparatus. In the Heisenberg picture, it maps classical effects $f: \Omega \rightarrow \mathbb [0,1]$ to
\[
\qcmap_E^\dagger (f) = f({\textrm true})\,E + f({\textrm false}) (\one - E).
\]
\end{ex}



When the system is quantum: $\mathcal A = \mathcal B(\hilbert)$, we call a classical-valued channel a quantum-to-classical channel, or QC-channel.
As an example we will determine the general structure of such QC-channels in the case where $\hilbert$ is finite dimensional and $\Omega$ finite.
This allows us to represent it as the CPTP map 
\[
\qcmap:\mathcal{B}(\hilbert)\rightarrow \funs(\Omega),
\]
From linearity of $\qcmap$ we can infer:
\[
\qcmap(\rho)=\{f_i(\rho)\}_{i \in \Omega} \equiv \sum_i f_i(\rho) \proj i \quad \text{with} \quad f_i:\mathcal{B}(\hilbert)\overset{lin}{\rightarrow} \mathbb{C}. 
\]
Theorem~\ref{th:riesz} tells us that there exists $A_i\in\mathcal{B}(\hilbert)$ such that $f_i(\rho)=\Tr(\rho A_i)$. The positivity of the map $\qcmap$ requires that for all $\rho$, $\Tr(A_i\rho)\geq 0$ which implies that $A_i\geq 0$ for all $i$. Moreover, the trace-preserving property of $\qcmap$ says that for all operator $X$,
\[
\Tr(\qcmap(Y))=\sum_i \Tr(A_i Y)=\Tr(Y)=\Tr(\one Y).
\]
Using in particular $Y = \ketbra i j$, this equivalently implies 
\[
\sum_i A_i = \one .
\]

Hence, the fact that the QC-channel $\qcmap$ is linear, trace-preserving and positive implies that it is represented by a set of operators $A_i \ge 0$, $i \in \Omega$, such that $\sum_i A_i = \one$ as
\begin{equation}
\label{qcmap}
\qcmap(\rho)=\bigl\{\Tr(A_i\rho)\bigr\}_{i \in \Omega} \equiv\sum_i\Tr(A_i\rho)\proj{i}.
\end{equation}
These operators $A_i$ are in fact \emph{effects} of the quantum system. Indeed, they automatically satisfy $0 \le A_i \le \one$, and are used to define the probabilities $\tr(\rho A_i)$ which make up the classical state $\qcmap(\rho)$.

Let us show that the complete positivity of $\qcmap$ does not impose any extra condition on these operators.
Any operator $W$ of the extended algebra $\mathcal B(\hilbert) \otimes \mathcal B(\hilbert')$ can be written as $W = \sum_j X_j \otimes Y_j$. The action of $\qcmap \otimes id$ on it is
\begin{align*}
(\qcmap\otimes id)(\Sigma_j X_j \otimes Y_j)&=\sum_{ij}\Tr(A_iX_j)\proj{i}\otimes Y_j=\sum_{ijk}\bra{k}\sqrt{A_i}X_j\sqrt{A_i}\ket{k}\proj{i}\otimes Y_j\\
    &=\sum_{ijk}\ket{i}\bra{k}\sqrt{A_i}X_j\sqrt{A_i}\ket{k}\bra{i}\otimes Y_j \\
	&=\sum_{ik}\underbrace{(\ket{i}\bra{k}\sqrt{A_i}\otimes\one )}_{E_{ik}}(\sum_j X_j\otimes Y_j)\underbrace{(\sqrt(A_i)\ket{k}\bra{i}\otimes \one )}_{E_{ik}^\dagger }
\end{align*}
Therefore, we find that $\qcmap \otimes id$ can be represented as $(\qcmap\otimes id)(W)=\sum_{ik}E_{ik}W E_{ik}^\dagger $, which is manifestly positive when $W$ is positive.

In summary, we find that a quantum-to-classical channel between finite systems $\qcmap:\mathcal{B}(\hilbert)\rightarrow \funs(\Omega)$ is defined, according to Equ.~\eqref{qcmap} by a set of quantum \emph{effects} $A_i$, $i \in \Omega$ such that $\sum_i A_i = \one$. 
A set of effects fulfilling this property is also a special instance of a \emph{positive operator-valued measure} (POVM), a terminology which will become more meaningful when we allow the target classical system to have continuously many pure states.


Consider for example the standard notion of an observable as hermitian operator $A=A^\dagger $. The spectral decomposition theorem tells us that we can write $A=\sum_i\alpha_i P_i$. Instead of labelling the spectral projectors $P_i$ by an integer $i$, we could also directly use the distinct eigenvalues $\alpha_i$ themselves as labels, and write
\[
A=\sum_{\alpha\in\Omega}\alpha P_\alpha,
\] 
where $\Omega=\{\alpha_1,...,\alpha_N\}$ and $P_{\alpha_i} \equiv P_i$. If the state is $\rho$, measuring the observable defined by $A$ puts the classical measurement apparatus in state $\alpha \in \Omega$ with probability
\[
p_\alpha = \Tr(\rho P_\alpha).
\]
Hence, the observable defines a map from the quantum state $\rho$ to a probabilistic state of the measurement apparatus with pure states $\Omega$. This is the QC-channel $\qcmap_A:\mathcal{B}(\hilbert)\rightarrow \funs(\Omega)$ with
\[
\qcmap_A(\rho) =  \bigl\{\Tr(\rho P_\alpha)\bigr\}_{\alpha\in\Omega} \equiv \sum_{\alpha \in \Omega} \tr(\rho P_\alpha) \proj \alpha,
\]
where the orthogonal vectors $\ket \alpha$ are used to label the classical states of the measurement apparatus. The set of effects defining this observables are simply the spectral projectors $P_\alpha$, and the eigenvalues correspond to the classical pure states $\Omega$.

We see that the old notion of observable defined by a self-adjoint operator corresponds only to a special type of QC-channel, or POVM, namely one whose effects are orthogonal \emph{projectors}. We say that it is a \emph{sharp}, or \emph{projective} observable. 

\begin{mydef}
We say that an observable represented by a QC-channel $\qcmap^\dagger $ is \emph{sharp} or \emph{projective} if it maps all sharp classical effects $E^2 = E$ to sharp quantum effects: $\qcmap^\dagger (E)^2 = \qcmap^\dagger (E)$.
\end{mydef}

Note that a complete set of projector are automatically mutually orthogonal, which is why orthogonality is not part of this definition:
\begin{lemma}
If $P_i$ are projectors such that $\sum_i P_i \le \one$, then $P_i P_j = 0$ for all $i \neq j$.
\end{lemma} 
\proof
For the proof, we just need to show that that it holds for two projectors $P$ and $Q$ such that $P+Q \le \one$. Multiplying the last inequality by $P$ from both sides we obtain $P + PQP \le P$ which implies $PQP \le 0$. But $PQP \ge 0$ by construction, so $PQP = 0$. This implies that for all states $\ket \psi$, $\bra \psi PQQP \ket \psi = 0$. Therefore, $\|QP \ket \psi\| =0$ for all $\ket \psi$, which implies $QP = 0$.
\qed

Despite this fact, the above concept of sharp observable is already more general than that associated with self-adjoint operators, in a way which may seem cosmetic at first, but which is important to realize. According to our definition, the set of value $\Omega$ can be \emph{any} measurable set, such as, for instance $\Omega = \mathbb R^n$, something which would require $n$ commuting self-adjoint operators to describe. 

\subsubsection{Coarse-graining of observables}

There is a natural way in which an observable can be ``weakened'', or made less sharp. Consider an observable represented by the QC channel $\qcmap : \mathcal{B}(\hilbert)\rightarrow \funs(\Omega)$ (Schr\"odiner picture). One may perform a measurement of this observable on a state $\rho$, obtaining the classical probability distribution $\qcmap(\rho)$, and then discard information by ``scrambling'' the output with a classical channel $\pi : \funs(\Omega)\rightarrow \funs(\Omega')$. Effectively, this is equivalent to having measured the new observable 
\[
\pi \circ \qcmap:  \mathcal{B}(\hilbert)\rightarrow \funs(\Omega'),
\]
where $\circ$ denotes the composition of maps: $(\pi \circ \qcmap)(\rho) = \pi(\qcmap(\rho))$.
We may say that this new observable is a \emph{coarse-graining} of $\qcmap$. It extracts less information about the system than $\qcmap$, for any reasonable concept of information. 

For instance, suppose $\qcmap$ is a sharp observable: $\qcmap(\rho)=\sum_{\alpha}\Tr(\rho P_\alpha)\proj{\alpha}$, where $P_\alpha$ are projectors, and we represent $\pi$ by the stochastic matrix $\pi_{\alpha\beta}$ as $\pi(\{p_\beta\}_{\alpha\in\Omega})=\{\sum_{\alpha}\pi_{\beta \alpha}q_\alpha\}_{\beta\in\Omega'}$. Then the coarse-grained observable $\pi \circ\qcmap$ is not in general a sharp observable anymore:
\[
(\pi \circ\qcmap)(\rho)= \bigl\{ \Tr(\rho E_\beta) \bigr\}_{\beta \in \Omega'} \ \text{with} \ E_\beta=\sum_\alpha \pi_{\beta \alpha}P_\alpha.
\]

\subsubsection{Accessible observables}
\label{sec:simpledecoh}

In order to gain some practice with these notions, imagine the following situation: we have two quantum systems $A$ and $B$ represented by the Hilbert spaces $\hilbert_A$ and $\hilbert_B$. There is some physical process carrying information from $A$ to $B$, represented by a CPTP map $\mathcal E : \mathcal{B}(\hilbert_A)\rightarrow\mathcal{B}(\hilbert_B)$. We would like to be able to gain information about system $A$, however we are just able to make measurements on system $B$. Suppose that we make a measurement represented by the QC map $\qcmap :  \mathcal{B}(\hilbert_B)\rightarrow \funs(\Omega)$. By making this measurement on system $B$, we are effectively measuring the observable
\[
\qcmap\circ\mathcal E \ : \ \mathcal{B}(\hilbert_A)\xrightarrow{\mathcal E}\mathcal{B}(\hilbert_B)\xrightarrow{\qcmap}\funs(\Omega)
\]
on system $A$. 

Hence, measuring $\qcmap$ on $B$ is perfectly equivalent, in terms of the classical information collected, to measuring $\qcmap\circ\mathcal E$ on system $A$ directly. This is just the Heisenberg picture: the channel represented in the Schr\"odinger picture by the CPTP map $\mc E$ sends the observable $\qcmap$ to $\qcmap\circ\mathcal E$. Therefore, we ought to be able to write this action directly in terms of the unital CP map $\mc E^\dagger$. Indeed, this is how the effects defining $\qcmap$ transform: if $X(\rho)=\{\Tr(X_{\alpha}\rho)\}_{\alpha\in\Omega}$, we have
\[
(\qcmap\circ\mathcal E)(\rho)=\bigl\{\Tr(X_{\alpha}\mathcal E(\rho))\bigr\}_{\alpha\in\Omega}=\bigl\{\Tr({\mathcal E^\dagger  (X_{\alpha})}
)\bigr\}_{\alpha\in\Omega}.
\]
Hence, $\mathcal E^\dagger$ maps the POVM elements of $\qcmap$ to those of $\qcmap \circ \mc E$.

In the sense just explained, we say that the channel $\mathcal E$ limits our ability to observe system $A$, and hence the information that we have about $A$. Specifically, it acts as a constraint of the observables of $A$, since it only allows us to measure effects of the form $\mathcal E^\dagger(X)$, where $X$ is any effect of system $B$.

As an example 
consider the channel $\mathcal E:\mathcal{B}(\hilbert_A)\rightarrow\mathcal{B}(\hilbert_B)$ with
\begin{equation}\label{eq:qc_channel}
\mathcal E(\rho)=\sum_i\Tr(\rho P_i)\proj{i},
\end{equation}
where $P_iP_j=\delta_{ij}P_i$. The adjoint map $\mc E^\dagger$ can be found by considering
\[
\tr(X \mc E(\rho)) = \sum_i \Tr(\rho P_i)\tr( X \proj{i} ) = \sum_i \Tr(\rho \bra i X \ket i P_i) = \tr(\rho \mc E^\dagger (X)),
\]
which yields
\[
\mc E^\dagger (X) = \sum_i \bra i X \ket i P_i.
\]
Therefore, if we only have access to system $B$, the only effects of system $A$ that we can indirectly observe are of the form $\sum_i x_i P_i$, where $x_i \in [0,1]$, which we recognise as the effects of the \emph{commutative} algebra
${\textrm alg}(P_1,...,P_n)$, which is isomorphic to the classical algebra $L(\{1,...,N\})$. Therefore, from the point of view of an observer of system $B$, system $A$ looks purely classical!  This specific channel $\mc E$ is a simple prototype of the phenomenon of \emph{decoherence} by which a quantum system appears classical.

\subsection{Stinespring dilation}
\label{sec:stinespring}

In the previous sections we have defined the concept of channel abstractly, as the most general type of map compatible with the state/effect structure. But we have not explained how such maps can occur in practical situations. In this section, we show that channels naturally occur in the description of the dynamics of \emph{open quantum systems}. More importantly, we show that \emph{any} channel has an appropriately unique such representation, which allows for a very powerful characterisation of quantum channels which has many consequences. The most general form of this result, the Stinespring dilation theorem, is arguably the most important mathematical result in quantum information theory.

We start by explaining how a simple decoherence channel such as the one mentioned in Section~\ref{sec:simpledecoh} can occur in a physically realistic situation. 
We consider two systems: the ``system of interest'', $S$, a two-dimensional quantum system with Hilbert space $\mc H_S = \mathbb C^2$, and an ``environment'' $E$ which is also quantum and two-dimensional, with Hilbert space $\mc H_E = \mathbb C^2$.
We imagine that they interact for some fixed amount of time, i.e., that they evolve under a common Hamiltonian. The result of this evolution is represented by a unitary map 
\[
U: \mc H_S \otimes \mc H_E \rightarrow \mc H_S \otimes \mc H_E.
\]
By choosing the interaction Hamiltonian, we can obtain any arbitrary unitary map $U$. For this example we consider
\[
U=\proj{0}_S\otimes\one _E+\proj{1}_S\otimes\bigl(\begin{smallmatrix}0&1\\ 1&0\end{smallmatrix} \bigr)_E.
\]
This unitary is also called the controlled NOT gate 
in the context of quantum computation, where it serves as a basic building block to construct more complex unitaries (See Ref.~\cite{nielsen}). 


Here the system $S$ can be thought of as a control bit, since if it starts in state $\ket 0$, then nothing happen to $E$, whereas, if it starts in state $\ket 1$, the unitary $\bigl(\begin{smallmatrix}0&1\\ 1&0\end{smallmatrix} \bigr)$ is applied to $E$, which swaps states $\ket 0$ and $\ket 1$ (a NOT operation in classical logic). Of course, this interpretation of the map makes sense only in this particular basis. For instance, in the basis $\ket + \propto \ket 0 + \ket 1$ and $\ket - \propto \ket 0 - \ket 1$, the role of control and target systems are exchanged. 

But let's keep with the original basis, and ask what happens if it is not the control qubit $S$ which starts in state $\ket 0$, but the target $E$ instead. Fixing the input state of one of the system amounts to considering the linear map
\begin{align*}
V:\hilbert_S &\rightarrow \hilbert_S\otimes\hilbert_E\\
\ket{\psi} &\mapsto U(\ket{\psi}\otimes \ket{0})
\end{align*}
Explicitely, we find
\[
V=\proj{0}_S\otimes\ket{0}_E+\proj{1}_S\otimes\ket{1}_E \equiv \ket{00}\bra{0} + \ket{11}\bra{1}.
\]
Such a map is a called an \emph{isometry}, in that, like a unitary map it satisfies $V^\dagger V = \one$, but it is not invertible, and $V V^\dagger$ is not the identity (but it is always a projector). It simply embeds a small Hilbert space into a bigger one while preserving the orthogonality of vectors. In this example; it represents $\mc H_S$ as the subspace of $\mc H_S \otimes \mc H_E$ spanned by $\ket{00}$ and $\ket{11}$. 

Moreover, if the initial state of $S$ is either $\ket 0$ or $\ket 1$, the map $V$ simply makes a \emph{copy} of the classical bit encoding this information. Again, this interpretation does not work in a different basis, but it indicates that \emph{some} information is being transmitted from $S$ to $E$. 

Finally, suppose that we only care about system $S$ after this interaction, i.e., we promise to only make further measurements on system $S$ only. In other words we discard $E$. In the density matrix formalism, this amounts to looking at the reduced state of system $S$ in the state after the interaction. If the original state of $S$ was represented by $\rho$, then after the interaction, the state of the joint system is $V \rho V^\dagger$. Discarding system $E$ yields the final state $\tr_E V \rho V^\dagger$ of the system $S$.

Altogether, this amounts to the map 
\[
\rho \mapsto \Tr_E V\rho V^\dagger, 
\]
which is a channel from $S$ to $S$. Explicitely,
\begin{align*}
\Tr_E V\rho V^\dagger  &=\sum_i(\bra{i}\otimes\one )V\rho_AV^\dagger  (\ket{i}\otimes \one )\\
                &=\sum_{ijk}(\bra{i}\otimes\one )(\proj{j}\otimes\ket{j})\rho(\proj{k}\otimes\bra{k})(\ket{i}\otimes \one )\\
                &=\sum_{ijk}\braket ij \braket ki \ket{j} \bra j\rho \ket k  \bra k
                =\sum_i\ket{i}\bra{i}\rho\ket{i}\bra{i} \equiv \sum_i \Tr\bigl( \rho \proj{i}\bigr)\proj{i},
\end{align*}
which is of the same form as the channel defined in Equ.\eqref{eq:qc_channel}. 

We see that the unitary evolution transforms a ``natural'' type of limitation (access only to a subsystem) to that of having access only to a commutative subalgebra of observables, as explained in Section~\ref{sec:simpledecoh} for this type of channel. The fact that only classical information is left in the system after this interaction is crucially linked to the fact that information about a basis was copied between $S$ and $E$ by $V$, as will be shown in a later part of this lecture.


In fact, a central result in quantum information theory is that any channel can occur in this way. This is stated by the following theorem, which have have here specialised to the finite-dimensional case in order to make the proof more accessible:
\begin{theo}[Stinespring]\label{th:stinespring}
A linear map $\channel:\mathcal{B}(\hilbert_A)\rightarrow\mathcal{B}(\hilbert_B)$ is completely positive if and only if there exists a Hilbert space $\hilbert_E$ (environment) and a linear map $V:\hilbert_A\rightarrow\hilbert_B\otimes\hilbert_E$ such that, for all state $\rho$,
\[
\channel(\rho)=\Tr_E(V\rho V^\dagger ),
\]
or, equivalently,
\[
\channel^\dagger  (X)= V^\dagger  (X\otimes\one _E)V.
\]
The map $\mc E$ is also trace-preserving, and hence a channel, if and only if $V^\dagger V = \one$. 
\end{theo}
\proof
This theorem in fact holds when the Hilbert spaces $\hilbert_A$ and $\hilbert_B$ are infinite-dimensional, but for simplicity we only prove it here in the finite-dimensional case, where we present a proof adapted from Ref.~\cite{choi}.
We immediately see that any map of the form $\channel(\rho) = \tr_E(V \rho V^\dagger)$ is completely positive because it is the composition of two completely positive operations, namely $\rho \mapsto V \rho V^\dagger$ and $\tr_E$. 


Conversely, suppose $\channel$ is completely positive. 
Let us define the matrix
\[
X_\channel := (\channel\otimes id)(\proj{\Omega}) \quad \text{with} \quad \ket{\Omega}=\frac{1}{d}\sum_i \ket{i}\otimes\ket{i},
\]
where $\ket i$ forms a basis of $\hilbert_A$.
Clearly, $X_\channel \ge 0$ since $\channel$ is CP and $\proj \Omega \ge 0$. 
This object $X_\channel$ is commonly referred to as the \emph{Choi matrix} of $\channel$. 
The good thing about the Choi matrix is that we can reconstruct the CP map $\channel$ from it as follows:
\begin{equation}
\label{eq:cj}
\begin{split}
\channel(\rho) &= (\one _1\otimes \bra{\Omega}_{23})(\channel_1\otimes id_2\otimes id_3)(\proj{\Omega}_{12}\otimes \rho_3)(\one _1 \otimes \ket{\Omega}_{23})\\
&= (\one _1\otimes\bra{\Omega}_{23})(X_{\channel}\otimes\rho_3)(\one _1\otimes\ket{\Omega}_{23})
\end{split}
\end{equation}
where we are now using three different copies of $\mc H_A$ which we labelled by the subscripts $1,2,3$ for clarity. This equation can be verified by direct expansion of the definition of the maximally entangled state $\ket \Omega$. 
This invertible map between CP maps and positive matrices is known as the \emph{Choi-Jamiolkowski} isomorphism.
The particular representation of the CP map that we are looking for simply amounts to diagonalising the Choi matrix. Indeed, since $X_\channel \ge 0$, we can write $X_\channel$ in diagonal form as $X_\channel = \sum_{i=1}^d \alpha_i \proj{\psi_i}$ where $\alpha_i \ge 0$ and $\braket {\psi_i}{\psi_j} = \delta_{ij}$. Substituting this diagonal form in Equ.~\eqref{eq:cj} yields the expression
\begin{align*}
\channel(\rho)&=(\one \otimes\bra{\Omega})(X_{\channel}\otimes\rho)(\one \otimes\ket{\Omega})\\
							&=\sum_{ijk}\alpha_i(\one \otimes\bra{k})\ket{\psi_i}\bra{k}\, \rho \, \ket{j}\bra{\psi_i}(\one \otimes\ket{j})
							= \sum_i E_i \rho E_i^\dagger,
\end{align*} 
where 
\[
E_i = \sum_k \sqrt{\alpha_i}(\one \otimes\bra{k})\ket{\psi_i}\bra{k}
\]
are called \emph{Kraus operators} for the CP map. They are directly related to the operator $V$ that we are looking for. 
Indeed, if we now take $\hilbert_E$ to be a hilbert space of dimension equal to the rank $d$ of $X_\channel$, and  $\{\ket{i}\}_{i=1}^{d}$ any orthonormal basis of $\hilbert_E$. Then if we define
\[
V := \sum_i E_i\otimes \ket{i}_E. 
\] 
We obtain 
\begin{align*}
\Tr_E V\rho V^\dagger 
&=\sum_i E_i\rho E_i^\dagger = \channel(\rho),
\end{align*}
which is the form that we were looking for. 
Moreover, if $\channel$ is also trace-preserving, then $\channel^\dagger$ is unital, which means that
\(
\channel^\dagger(\one) = V^\dagger V = \one. 
\)  
\qed

The isometry $V$ from Theorem~\ref{th:stinespring}, together with the Hilbert space $\hilbert_E$, is called a Stinespring dilation of $\channel$. An important aspect of this dilation is that it is unique up to a \emph{partial isometry} on $\hilbert_E$. A partial isometry $W$ preserves the scalar product on its range, but can send some vectors to zero. It is characterised by the fact that $W^\dagger W$ is a projector, which also implies that $W W^\dagger$ is a projector.
\begin{theo}[Stinespring, uniqueness]\label{th:stinespringunique}
Let $\hilbert_{E_1}$ and $\hilbert_{E_2}$ and $V_1 : \hilbert_A \rightarrow \hilbert_B \otimes \hilbert_{E_1}$, $V_2 : \hilbert_A \rightarrow \hilbert_B \otimes \hilbert_{E_2}$ be such that
\[
V_1^\dagger (X \otimes \one_{E_1}) V_1 = V_2^\dagger (X \otimes \one_{E_2}) V_2
\]
for all $X \in \mathcal B(\hilbert_B)$. Then
there is a partial isometry $W:  \hilbert_{E_1} \rightarrow \hilbert_{E_2}$, such that
\[ 
V_2 = (\one_B \otimes W) V_1 .
\]
\end{theo} 
\proof
Consider the span $\mathcal K$ of all vectors of the form $(X \otimes \one_{E_1}) V_1 \ket \psi$, for arbitrary state $\ket \psi$ and operator $X$. Let us show that there is a well-defined linear map $\tilde W$ on $\mathcal K$ sending an arbitrary element of the form $\sum_i c_i (X_i \otimes \one_{E_1}) V_1 \ket {\psi_i}$ to $\sum_i c_i (X_i \otimes \one_{E_2}) V_2 \ket {\psi_i}$ (with just $V_1$ replaced by $V_2$).  Suppose $\sum_i c_i (X_i \otimes \one_{E_1}) V_1 \ket {\psi_i} = \sum_i d_i (Y_i \otimes \one_{E_1}) V_1 \ket {\phi_i}$, then we have to show that the images are also equal. The norm of the difference between these two different representations of the vector (which equals zero) is a linear combination of components of operators of the form $V_1^\dagger (A \otimes \one) V_1$. Since in each term $V_1$ can be replaced by $V_2$---because the two isometries represent the same channel---then we obtain that the norm of $\sum_i c_i (X_i \otimes \one_{E_2}) V_2 \ket {\psi_i} - \sum_i d_i (Y_i \otimes \one_{E_2}) V_2 \ket {\phi_i}$ is zero as well: those two vectors are equal. Therefore the map $\tilde W$ is well-defined on $\mathcal K$, and linear by construction.


We also extend the definition of $\tilde W$ to the whole of $\hilbert_B \otimes \hilbert_{E_1}$, by requiring that it sends any vector $\ket \phi$ in the orthogonal complement of $\mathcal K$ to zero.

It follows from this definition that $\tilde W (X \otimes \one) V_1 = \tilde W (X \otimes \one) V_2$ for all $X$,
and in particular,
\[
\tilde W V_1 = V_2.
\]
The operator $\tilde W$ is a partial isometry because, due to the fact that both dilations represent the same channel,
\[
\begin{split}
\bra \psi V_2^\dagger (X^\dagger \otimes \one) (Y \otimes \one) V_2 \ket \psi  
&= \bra \psi V_2^\dagger \bigl( X^\dagger Y \otimes \one \bigr) V_2 \ket \psi  \\
&= \bra \psi V_1^\dagger \bigl( X^\dagger Y \otimes \one \bigr) V_1 \ket \psi 
= \bra \psi V_1^\dagger (X^\dagger \otimes \one) (Y \otimes \one) V_1 \ket \psi 
\end{split}
\]
for all $X$, $Y$ and all $\ket \psi$.
Moreover, for all $X$, $Y$,
\[
\tilde W (X \otimes \one) (Y \otimes \one) V_1 = (X \otimes \one) (Y \otimes \one) V_2 = (X \otimes \one) \tilde W (Y \otimes \one) V_1,
\]
which shows that $\tilde W (X \otimes \one) \ket \phi  = (X \otimes \one) \tilde W \ket \phi$ for all $\ket \phi \in \mathcal K$. 
Moreover, if $\ket {\phi'}$ belongs to the orthogonal complement $\mathcal K^\perp$, we have $\tilde W \ket{\phi'} = 0$ by definition of $\tilde W$, but also, since $\tilde W^\dagger$ maps onto $\mathcal K$, then for all $\ket \psi$ and all $X$, $\bra \psi \tilde W (X \otimes \one) \ket{\phi'} = 0$, which implies $ \tilde W (X \otimes \one) \ket{\phi'} = 0$. Therefore, $ \tilde W (X \otimes \one) \ket{\phi'} = (X \otimes \one) \tilde W \ket{\phi'}$ also for all $\phi' \in \mathcal K^\perp$. We conclude that 
\[
\tilde W (X \otimes \one) = (X \otimes \one) \tilde W
\]
for all $X$, which implies
\[
\tilde W = \one \otimes W
\]
for some operator $W$ which inherits its isometric property from $\tilde W$.
\qed



